\documentclass{aa}  

\usepackage{graphicx}
\usepackage{txfonts}
\usepackage{xspace}
\usepackage{xcolor}
\usepackage{bm}
\usepackage[colorlinks,linkcolor=red,citecolor=blue,urlcolor=blue ]{hyperref}
\usepackage{mathptmx} 
\usepackage{booktabs}

\newcommand{\y}{\ensuremath{y}\xspace}
\newcommand{\KiDS}{KiDS-1000\xspace}
\newcommand{\Planck}{\textit{Planck}\xspace}
\newcommand{\milca}{{\sc milca}\xspace}
\newcommand{\nilc}{{\sc nilc}\xspace}
\newcommand{\hmx}{\textsc{HMx}\xspace}
\newcommand{\bahamas}{\textsc{bahamas}\xspace}
\newcommand{\hmcode}{\textsc{HMCode}\xspace}

\newcommand{\nbody}{$N$-body\xspace}
\newcommand{\om}{\ensuremath{\Omega_\mathrm{m}}\xspace}
\newcommand{\logtheat}{\ensuremath{\log_{10}{\left(\frac{T_{\rm AGN}}{\mathrm{K}}\right)}}\xspace}
\newcommand{\logtheatinline}{\ensuremath{\log_{10}{\left(T_{\rm AGN}/\mathrm{K}\right)}}\xspace}
\newcommand{\Sigmaalpha}{\ensuremath{\Sigma_8^{0.2}}\xspace}
\newcommand{\software}{\textsc}

\newcommand{\pCl}{pseudo-$C_\ell$\xspace}
\newcommand{\alphaCIB}{\ensuremath{\alpha_\mathrm{CIB}}\xspace}

\newcommand{\eg}{e.g.\xspace}

\newcommand{\iMpc}{\,h\mathrm{Mpc}^{-1}}

\usepackage{environ}
\NewEnviron{splitequation}{%
    \begin{equation}\begin{split}
        \BODY
    \end{split}\end{equation}}

\newcommand{\diff}{\mathrm{d}}
\DeclareRobustCommand{\vec}[1]{\bm{#1}}
\newcommand{\mtrx}[1]{\mathsf{#1}}
\newcommand{\Cov}[1]{\ensuremath{\mathrm{Cov}\left[#1\right]}}

\usepackage{listings}

\definecolor{codegreen}{rgb}{0,0.6,0}
\definecolor{codegray}{rgb}{0.5,0.5,0.5}
\definecolor{codepurple}{rgb}{0.58,0,0.82}
\definecolor{backcolour}{rgb}{0.95,0.95,0.92}

\lstdefinestyle{codestyle}{
    backgroundcolor=\color{backcolour},   
    commentstyle=\color{codegreen},
    keywordstyle=\color{magenta},
    numberstyle=\tiny\color{codegray},
    stringstyle=\color{codepurple},
    basicstyle=\ttfamily\footnotesize,
    breakatwhitespace=false,         
    breaklines=true,                 
    captionpos=b,                    
    keepspaces=true,                 
    numbersep=5pt,                  
    showspaces=false,                
    showstringspaces=false,
    showtabs=false,                  
    tabsize=2
}
\newcommand{\chisqCS}{\ensuremath{150.7}\xspace}
\newcommand{\ndofCS}{\ensuremath{117.1}\xspace}
\newcommand{\chisqXcorr}{\ensuremath{28.0}\xspace}
\newcommand{\ndofXcorr}{\ensuremath{36.2}\xspace}
\newcommand{\chisqJoint}{\ensuremath{179.3}\xspace}
\newcommand{\ndofJoint}{\ensuremath{155.4}\xspace}

\newcommand{\omJoint}{\ensuremath{0.342^{+0.042}_{-0.037}}\xspace}
\newcommand{\sigmaeightJoint}{\ensuremath{0.692^{+0.05}_{-0.039}}\xspace}
\newcommand{\SeightJoint}{\ensuremath{0.751^{+0.02}_{-0.017}}\xspace}
\newcommand{\logtheatJoint}{\ensuremath{7.96^{+0.2}_{-0.48}}\xspace}

\newcommand{\SigmaalphaXcorr}{\ensuremath{0.723^{+0.046}_{-0.032}}\xspace}

\newcommand{\SeightJointImprovement}{\ensuremath{40\,\%}\xspace}
\newcommand{\SigmaalphaJointImprovement}{\ensuremath{15\,\%}\xspace}

\newcommand{\chisqBB}{\ensuremath{128.3}\xspace}
\newcommand{\pteBB}{\ensuremath{0.29}\xspace}
\newcommand{\chisqTBMILCA}{\ensuremath{36.5}\xspace}
\newcommand{\pteTBMILCA}{\ensuremath{0.63}\xspace}

\newcommand{\SNEE}{\ensuremath{20.9\,\sigma}\xspace}
\newcommand{\SNTEMILCA}{\ensuremath{9.4\,\sigma}\xspace}

\usepackage{CJKutf8}

\begin{document}

\title{Joint constraints on cosmology and the impact of baryon feedback: Combining KiDS-1000 lensing with the thermal Sunyaev-Zeldovich effect from \textit{Planck} and ACT}

    \titlerunning{Cosmology from KiDS-1000 x tSZ}
    
    \author{Tilman~Tröster
          \inst{1}\fnmsep\thanks{Email: ttr@roe.ac.uk}
          \and
          Alexander~J.~Mead\inst{1}
          \and
          Catherine Heymans\inst{1,2}
   \and Ziang Yan \protect\begin{CJK*}{UTF8}{gbsn}(颜子昂)\protect\end{CJK*}\inst{3}
   \and David Alonso\inst{4}
   \and Marika Asgari\inst{1} 
   \and Maciej Bilicki\inst{5}
   \and Andrej Dvornik\inst{2}
   \and Hendrik Hildebrandt\inst{2}
   \and Benjamin Joachimi\inst{6}
   \and Arun Kannawadi\inst{7}
   \and Konrad Kuijken\inst{8}
   \and Peter Schneider\inst{9}
   \and HuanYuan Shan \protect\begin{CJK*}{UTF8}{gbsn}(陕欢源)\protect\end{CJK*}\inst{10,11}
   \and Ludovic van Waerbeke\inst{3}
   \and Angus~H.~Wright\inst{2}
          }

   \institute{Institute for Astronomy, University of Edinburgh, Royal Observatory, Blackford Hill, Edinburgh, EH9 3HJ, UK 
   \and
   Ruhr University Bochum, Faculty of Physics and Astronomy, Astronomical Institute (AIRUB), German Centre for Cosmological Lensing, 44780 Bochum, Germany
   \and
   Department of Physics and Astronomy, University of British Columbia, 6224 Agricultural Road, Vancouver, BC, V6T 1Z1, Canada
   \and
   Department of Physics, University of Oxford, Denys Wilkinson Building, Keble Road, Oxford OX1 3RH, United Kingdom
   \and
   Center for Theoretical Physics, Polish Academy of Sciences, al. Lotników 32/46, 02-668 Warsaw, Poland
   \and
   Department of Physics and Astronomy, University College London, Gower Street, London WC1E 6BT, UK
   \and
   Department of Astrophysical Sciences, Princeton University, 4 Ivy Lane, Princeton, NJ 08544, USA
   \and
   Leiden Observatory, Leiden University, P.O.Box 9513, 2300RA Leiden, The Netherlands
   \and
   Argelander-Institut f. Astronomie, Univ. Bonn, Auf dem Huegel 71, D-53121 Bonn, Germany
   \and
   Shanghai Astronomical Observatory (SHAO), Nandan Road 80, Shanghai 200030, China
   \and
   University of Chinese Academy of Sciences, Beijing 100049, China
}

\date{}

\abstract{
We conduct a \pCl analysis of the tomographic cross-correlation between $1000\,\mathrm{deg}^2$ of weak-lensing data from the Kilo-Degree Survey (KiDS-1000) and the thermal Sunyaev-Zeldovich (tSZ) effect measured by \Planck and the Atacama Cosmology Telescope (ACT). 
Using \hmx, a halo-model-based approach that consistently models the gas, star, and dark matter components, we are able to derive constraints on both cosmology and baryon feedback for the first time from these data, marginalising over redshift uncertainties, intrinsic alignment of galaxies, and contamination by the cosmic infrared background (CIB). 
We find our results to be insensitive to the CIB, while intrinsic alignment provides a small but significant contribution to the lensing--tSZ cross-correlation. 
The cosmological constraints are consistent with those of other low-redshift probes and prefer strong baryon feedback. 
The inferred amplitude of the lensing--tSZ cross-correlation signal, which scales as $\sigma_8\left(\om/0.3\right)^{0.2}$, is low by $\sim 2\,\sigma$ compared to the primary cosmic microwave background constraints by \Planck. 
The lensing--tSZ measurements are then combined with \pCl measurements of KiDS-1000 cosmic shear into a novel joint analysis, accounting for the full cross-covariance between the probes, providing tight cosmological constraints by breaking parameter degeneracies inherent to both probes. 
The joint analysis gives an improvement of \SeightJointImprovement on the constraint of $S_8=\sigma_8\sqrt{\om/0.3}$ over cosmic shear alone, while providing constraints on baryon feedback consistent with hydrodynamical simulations, demonstrating the potential of such joint analyses with baryonic tracers such as the tSZ effect.
We discuss remaining modelling challenges that need to be addressed if these baryonic probes are to be included in future precision-cosmology analyses. 
}


\maketitle
%
\section{Introduction}

Gravitational lensing by large-scale structure induces an angular correlation of galaxy shapes on the sky, which in principle contains a wealth of information about the constituents of the Universe. 
The signal from this `weak' gravitational lensing is determined by both the Universe's expansion history and the growth of perturbations. 
Exquisite measurements of hundreds of millions of galaxy shapes that are expected to be made in the coming decade by galaxy surveys, such as the Legacy Survey of Space and Time (LSST) of the Vera C. Rubin Observatory \citep{Ivezic2019-Rubin}, the \textit{Euclid} Mission \citep{Euclid-RedBook}, and the Nancy Grace Roman Space Telescope \citep{Spergel2015-NancyRoman}, will allow for the impact of dark energy to be assessed with a precision that has never before been possible. 
Perhaps most excitingly, the manner of growth of large-scale structure is precisely determined by the underlying gravity theory, and thus general relativity will be robustly tested in a cosmological setting.

The details of how galaxy shapes are distorted is governed by the degree of clustering of large-scale structure between the observed galaxy and Earth. Unfortunately, those structures that contribute most to the correlation of galaxy shapes (over the range of scales where they are best measured) are dense `non-linear' structures whose evolution cannot be calculated via linear (or even quasi-linear) perturbation theory. Structure formation must therefore be simulated, and the clustering statistics extracted from the simulations, in order to compare models to observed data. The simplest clustering statistic is the two-point function, which can be precisely measured from simulations (\eg \citealt{Heitmann2010}; although see \citealt{Smith2014}). Emulators \citep[\eg][]{Lawrence2017, Angulo2021, Knabenhans2021-EuclidEmulator2} can then be constructed to interpolate intelligently between simulation outputs for different cosmological parameters, or physically motivated fitting formulae \citep[\eg][]{Takahashi2012, Mead2021-HMCode2020} can be developed.

If gravity were the only important force in the Universe, then the modelling of structure formation at the level of precision required for the measurements from forthcoming surveys would be a solved problem. 
Unfortunately, non-gravitational processes can have a significant impact on the distribution of matter. 
Gas cools, contracts, and forms stars, some of which eventually explode as energetic supernovae; black holes form, and complicated feedback mechanisms transfer gravitational energy from accretion events into the gas surrounding galaxies. 
In cosmology these processes are known collectively as `baryonic feedback', and the complex physics of how they operate is poorly understood in comparison with pure gravitational clustering. While it is possible to run accurate hydrodynamic simulations that incorporate gas physics \citep[\eg][]{Springel2010} such as cooling and radiating, the details of processes such as active galactic nucleus (AGN) and supernova feedback must be included schematically, via sub-grid recipes \citep[\eg][]{Schaye2010, McCarthy2017}, and it is difficult to assess the numerical convergence and accuracy of such simulations. 
The unknown impact of feedback on the matter distribution is one of the biggest uncertainties in current lensing analyses \citep[\eg][]{Chisari2019} and if not addressed will limit the ability of forthcoming surveys to test dark energy and gravity theories.
Attempts have been made to use these effects of baryon feedback on lensing observables as a probe of galaxy formation  \citep[\eg][]{Harnois-Deraps2015,Foreman2016,MacCrann2017, Huang2021,Yoon2021}, but the inferred constraints on baryon feedback have been weak for past and current surveys.

The thermal Sunyaev-Zeldovich \citep[tSZ;][]{Sunyaev1972} effect arises when cosmic microwave background (CMB) photons are Compton-scattered off free electrons, predominantly hot electrons in galaxy clusters. 
This results in a spectral distortion of the CMB black-body spectrum with a characteristic frequency dependence. 
Maps of this Compton-\y signal can be extracted from multi-frequency CMB temperature data \citep[\eg][]{Planck2015-tSZ}. 
The tSZ effect therefore traces hot ionised gas in the Universe, which is itself correlated with large-scale structure. 
This correlation contains useful cosmological information \citep[\eg][]{Planck2015-tSZ, Horowitz2017, Bolliet2018, Makiya2020, Mitchell2021} and exhibits a particularly strong scaling with the amplitude of perturbations \citep{Refregier2002, Komatsu2002}. 
Of particular interest to weak-lensing cosmology is the fact that it is this same hot gas that has been redistributed by baryonic feedback, and therefore the tSZ effect provides an important external constraint on the magnitude of baryonic feedback.

The tSZ--galaxy cross-correlation has been investigated \citep[\eg][]{Vikram2017, Makiya2018, Pandey2019, Tanimura2020, Chiang2020, Koukoufilippas2020, Schaan2021, Amodeo2021, Yan2021}, but constraints on cosmological parameters are dependent on the unknown galaxy bias, which must either be taken from external data or else modelled in the data analysis. 
Stacking of tSZ observations of large-scale structure filaments \citep[\eg][]{Tanimura2019, de-Graaff2019} has shed light on the distribution of diffuse gas. 
In this paper we utilise measurements of weak lensing and the tSZ effect to place simultaneous constraints on the cosmological model and on the strength of baryonic feedback. 
The cross-correlation between lensing and the tSZ effect has been measured previously for both CMB lensing \citep[\eg][]{Hill2014} and for galaxy lensing \citep[\eg][]{Van-Waerbeke2014, Hojjati2017, Osato2020, Gatti2021, Pandey2021}. 
In such works, model parameters were either fixed or only varied within tight priors, with no coupling between parameters that describe the effect of baryon feedback on the lensing and the tSZ effect. 
Here we build on these previous works and analyse the tomographic cross-correlation between the tSZ effect and weak lensing from the $1000\,\mathrm{deg}^2$ release of the Kilo-Degree Survey (KiDS-1000) in a fully Bayesian manner, deriving posteriors on cosmological and baryon feedback parameters while marginalising over redshift uncertainties, the intrinsic alignment (IA) of galaxies, and contamination by the cosmic infrared background (CIB). 
We then combine the lensing--tSZ cross-correlation with KiDS-1000 cosmic shear in a novel joint analysis. 
By using the \citet{Mead2020} \hmx model, which consistently accounts for gas, stars, and dark matter in the whole joint-data vector, we are able to infer constraints on cosmology, as well as the `AGN temperature' parameter that governs the impact of feedback on both the lensing--tSZ cross-correlation and cosmic shear.

This paper is organised as follows: In Sect.~\ref{sec:data} we summarise the lensing and tSZ data, as well as the estimators and summary statistics used; in Sect.~\ref{sec:modelling} we present the modelling of the signal, its covariance, and systematics; Sect.~\ref{sec:results} presents the results, including measured data vectors and inferred cosmological constraints; and we summarise our findings in Sect.~\ref{sec:conclusions}.

\section{Data}
\label{sec:data}

\subsection{Kilo-Degree Survey}
We analysed the fourth data release of the Kilo-Degree Survey \citep{kuijken/etal:2019}, KiDS-1000, which spans $\sim 1000\,\mathrm{deg}^{2}$, split equally between an equatorial and southern imaging stripe. 
KiDS is unique in terms of its complete matched-depth nine-band imaging across the optical to the near-infrared, with relatively uniform and high-resolution seeing in the $r$-band. 
This combination of spatial and wavelength resolution provides an exquisite dataset for weak gravitational lensing studies, allowing for sensitive control of data-related systematics \citep{giblin/etal:2021, Hildebrandt2021-NZ}. 

Galaxy shear estimates are determined using the model-fitting {\it lens}fit method \citep{miller/etal:2013}, validated and calibrated using realistic KiDS image simulations \citep{kannawadi/etal:2019}. 
The percent-level uncertainty on this calibration correction is accounted for in our covariance calculation \citep[see Sect. 3.4 of][for details]{Joachimi2021}. 
Tomographic galaxy redshift distributions are determined using overlapping spectroscopic training data and a self-organising map to account for incompleteness in the training sample \citep{wright/etal:2020}.  
The accuracy of the resulting distributions are validated with a cross-correlation clustering analysis and mock survey analysis \citep{Hildebrandt2021-NZ}.  
The measured uncertainty from these tests is used as a prior for a set of correlated nuisance parameters that account for any residual correlated calibration errors in the mean redshift of each tomographic bin \citep[see Sect. 3.3 of][and Sect.~\ref{sec:photoz} for details]{Joachimi2021}.  
The full suite of KiDS-1000 data products, including imaging, catalogues, shear and redshift calibrations, is available to download\footnote{KiDS-1000 data access: \url{http://kids.strw.leidenuniv.nl/DR4/lensing.php}}.

\subsection{Compton-\y maps}
The construction of Compton-\y maps, which trace the tSZ effect, usually proceeds by taking linear combinations of intensity maps at different frequencies of CMB experiments, potentially with scale-dependent or spatially varying weights. 
\citet{Planck2015-tSZ} provide two such Compton-\y map reconstructions, using two different internal linear combination (ILC) methods:
\milca \citep[modified internal linear combination algorithm;][]{Hurier2013} and \nilc \citep[needlet independent linear combination; ][]{Remazeilles2011}. 
We used the \milca map as our fiducial \y-map in our analysis but confirmed that the results hold for the \nilc map as well. 
Furthermore, we checked our sensitivity to CIB contamination by using Compton-\y maps from \citet{Yan2019} where the CIB has been subtracted from the \Planck frequency maps. 
Assessment and mitigation of possible CIB contamination of our \y-maps is further discussed in Sect.~\ref{sec:CIB}.

We applied a Galactic mask that covers the 40\,\% of the sky with the strongest dust emissions, as well as a point-source mask, as provided by \citet{Planck2015-tSZ}. 
The extra-galactic lensing signal is assumed not to be correlated with emissions from the Milky Way, such that the Galactic mask primarily serves to reduce noise in the measurement. 
Furthermore, as the \KiDS footprint avoids the Galactic plane, the Galactic mask has only a small effect on the overlap between the shear and Compton-\y maps, such that the unmasked area of the overlap between the \KiDS and \Planck data makes up 85\,\% of the footprint of the \KiDS data. 
The maps are provided in \software{HEALPix} \citep{Gorski2005} format, with $\texttt{nside}=2048$, in Galactic coordinates. 

The \Planck-derived \y-maps have a resolution of $10\ \mathrm{arcmin}$, which limits the amount of small-scale information that can be extracted. 
In contrast, Compton-\y maps constructed from the combination of the fourth data release of the Atacama Cosmology Telescope \citep[ACT;][]{Mallaby-Kay2021} and \Planck data, benefit from a much smaller beam of $1.6\ \mathrm{arcmin}$ \citep{Madhavacheril2020}. 
Unlike the \Planck-derived maps, the ACT \y-maps only partially cover the northern \KiDS patch, while they do not overlap with the southern \KiDS patch at all. 
After applying the same 40\,\% Galactic and point source masks as for the \Planck-derived maps, the overlap between the ACT and \KiDS data is reduced to 30\,\% of the area covered by the \KiDS data, or 35\,\% of the area of the overlap between the \Planck-derived \y-maps and \KiDS. 

We transformed the ACT \y-maps into \software{HEALPix} format ($\texttt{nside}=2048$) using \software{pixell}\footnote{\url{https://github.com/simonsobs/pixell}}. 
Unlike the \Planck-derived maps, the ACT maps are provided in equatorial coordinates, the same as the KiDS-1000 data.
When measuring the shear--tSZ cross-correlations with the \Planck-derived maps, we used a shear catalogue that had been transformed into Galactic coordinates \citep[see \eg][]{Xia2020}, which avoids interpolation artefacts that would be introduced if the \y-map were transformed into a different coordinate system.

\subsection{Summary statistics and estimators}
\label{sec:summary-stats}
Our summary statistic are the angular (cross-)power spectra of the shear field and tSZ effect. 
We estimated the angular power spectra using the \pCl method, specifically the \software{MASTER} \citep{Hivon2002} estimator as implemented in the \software{NaMaster}\footnote{\url{https://github.com/LSSTDESC/NaMaster}} \citep{Alonso2019} software. 
The method proceeds by first naively estimating the angular power spectra of the masked maps. 
These estimates are biased and thus give the method its name: \pCl. 
The bias introduced by the masks can be calculated and is encoded in the so-called mode-mixing matrices. 
Given the mode-mixing matrices, an unbiased estimate of the (binned) angular power spectra can be obtained from the biased \pCl estimates. 
We refer the interested reader to \citet{Alonso2019}, who map these steps out in detail (see their Eqs. (7), (9), (D2), and (13) for the steps outlined above).

Estimating shear power spectra using \pCl methods comes with a number of subtleties due to the amount of small-scale power in the masks involved, which can affect the numerical stability of the algorithm. 
Here we closely followed the methodology of \citet{Nicola2021}. 
We pixelised the shear catalogues into \software{HEALPix} maps $\hat{\vec\gamma}$ with $\texttt{nside}=2048$:
\begin{equation}
    \label{equ:shear-map}
    \hat{\vec\gamma}(\vec n_i) = \frac{1}{w(\vec n_i)}\sum_{j\in\vec n_i} w_j \vec \epsilon_j\,,
\end{equation}
where $\vec n_i$ denotes the position of the $i$-th pixel and the sum runs over all galaxies in that pixel, which are characterised by their weight $w_j$ and the two ellipticity components, which we denote here with $\vec \epsilon_j$. 
In our case, the weights $w_j$ are given by the \textit{lens}fit weights and account for uncertainty in the shape measurement. 
The weight map $w(\vec n_i)$ is defined as
\begin{equation}
    \label{equ:weight-map}
    w(\vec n_i) = \sum_{j\in\vec n_i} w_j\,.
\end{equation}
The multiplicative shear bias was accounted for by scaling all ellipticity estimates in a tomographic bin by $\vec \epsilon\to \frac{1}{1+m_\mathrm{z}}\vec \epsilon$, where we used the multiplicative bias values $m_\mathrm{z}$ in \citet{giblin/etal:2021}. 
The additive shear bias was accounted for by subtracting the weighted ellipticity mean in each tomographic bin, following \citet{giblin/etal:2021}. 
The weight maps of Eq.~\eqref{equ:weight-map} take the role of the masks for the shear maps of the \pCl estimator. 
All cross-power spectra were estimated using their respective masks, that is, the measurements are not restricted to the overlap region. 

The auto-power spectra are subject to a noise bias, which we accounted for analytically, using the estimate \citep{Nicola2021}
\begin{equation}
    N_\ell = \Omega_\mathrm{pix}\left\langle \sum_{j\in \vec n_i} w_j^2\sigma_{\gamma,j}^2\right\rangle_\mathrm{pix}\,,
\end{equation}
where $\Omega_\mathrm{pix}$ is the pixel area, $\sigma_{\gamma,j}^2 = \frac{1}{2}|\vec \epsilon_j|^2$ with $|\vec \epsilon|^2 = \epsilon_1^2 +\epsilon_2^2$, and the average is taken over all pixels. 
The accuracy of this estimate was confirmed using a large number of Gaussian random field mocks with realistic noise and the real data weight maps of Eq.~\eqref{equ:weight-map}.

The angular power spectra were estimated in 12 logarithmically spaced bins between $\ell=51$ and $\ell=2952$. 
When discarding the lowest two and highest two $\ell$-bins, this binning scheme results in the same eight logarithmically spaced multipole bins between $\ell=100$ and $\ell=1500$ that were used in the KiDS-1000 analyses \citep{Joachimi2021, Asgari2021-CS, Heymans2021, Troester2021}. 
Since the KiDS-1000 cosmic shear methodology was validated only on the $\ell\in(100,1500)$ range, we restricted our analysis to the same $\ell$-range. 
Furthermore, this $\ell$-range captures most of the signal, with the signal-to-noise ratio increasing by only $\sim 10\,\%$ for the full $\ell\in(51,2952)$ range. 

We note that our \pCl approach differs from the KiDS-1000 cosmic shear band-power methodology of \citet{Asgari2021-CS} and from the recent \pCl methodology of \citet{Loureiro2021}.
Our analysis choice, which does not impact the KiDS-1000 cosmic shear-only constraints (see Appendix~\ref{app:bp-vs-pcl}), is motivated by the requirements of the lensing--tSZ cross-correlation analysis and support for the computation of the full spin-2 Gaussian data covariance by \software{NaMaster} (see Sect.~\ref{sec:covariance}). 
The angular power spectrum estimates were also checked against those estimated with \software{PolSpice}\footnote{\url{http://www2.iap.fr/users/hivon/software/PolSpice/}} \citep{Szapudi2001, Chon2004} and found to be in excellent agreement. 
The measurement scripts, associated tools, and the measurements themselves are being made publicly available\footnote{\url{https://github.com/tilmantroester/KiDS-1000xtSZ}}.

\section{Modelling}
\label{sec:modelling}
\subsection{Three-dimensional power spectra}
\label{sec:3d-model}

To model our data and extract meaningful cosmological constraints we followed \cite{Mead2020} and used the \hmx approach, which is designed to reproduce power spectra measured from the \bahamas \citep{McCarthy2017} suite of cosmological hydrodynamic simulations, which in turn are calibrated to reproduce the observed galaxy stellar mass function and baryon fraction in groups and clusters. 
\hmx works at the three-dimensional level, providing $P(k)$ rather than $C_\ell$, because once $P(k)$ is accurate the $C_\ell$ are automatically also accurate via the projection integrals; modelling and physical understanding are also more straightforward in three dimensions where the fields naturally live. 
The three-dimensional source of the tSZ effect is the electron pressure, while that of shear is the matter overdensity, and \hmx reproduces  the electron pressure--matter power spectrum from \bahamas at the $15$ percent level for $z<2$ and $k\lesssim7\iMpc$ while at the same time providing matter--matter power spectra that are accurate at the $2$ percent level\footnote{The estimate of the \hmx accuracy does not account for sampling variance, which is significant for the electron pressure for two reasons: first, unlike the matter power spectrum, sampling variance cancels to a much smaller degree when taking the ratio between baryonic and dark matter-only power spectra. Second, due to the sensitivity of the electron pressure power spectra to massive, and thus rare, haloes, sampling variance is still large for the 400$\iMpc$ \bahamas simulation boxes.}. 
Ultimately we wish to use \hmx to extract meaningful constraints from combined shear--shear and shear--tSZ data.

Underpinning \hmx is the halo model \cite[\eg][]{Cooray2002}, where halo density profiles are broken down into their constituent dark matter, gas and stellar components. The gas is then given thermal properties such that the halo electron pressure can be simultaneously calculated with the matter. 
This occurs both at the level of individual haloes and the expelled background gas, which is assumed to be heated by AGN activity. 
Fitting parameters of the model to data from \bahamas provides the link in \hmx between gas thermal properties and the density. 
Any standard halo model cross-correlation calculation \citep[\eg][]{Hill2014, Ma2015} provides a degree of cross-talk between the two fields through the common linear spectrum, halo mass function and bias, but most often the halo profiles of the two fields are taken to be independent. 
For example, to model the shear--tSZ signal, an \cite{Arnaud2010} profile may be adopted for the pressure while a (completely separate) \citeauthor*{Navarro1997} (\citeyear{Navarro1997}; NFW) profile may be adopted for the matter. 
The advantage of \hmx is that the halo profiles are also directly coupled to each other, which allows the electron pressure (which probes the hot gas) to directly affect the overall matter profile. 
The model parameters, consisting of the concentration of dark matter halo density profiles and its redshift evolution, the polytropic index of gas bound in haloes, the halo mass below which haloes have lost half their initial gas content, the ratio of halo temperature to that of the virial equilibrium, and the temperature of the warm-hot intergalactic medium with its redshift evolution, are fit to power spectra from simulations with different feedback strengths. 
All fitted parameters are then related to a single `AGN temperature' parameter, \logtheatinline, which governs the strength of feedback; increasing the temperature increases the tSZ signal while simultaneously decreasing the matter signal as gas is forced from the halo by the increased AGN activity. 
This is particularly helpful in gaining an understanding of the impact baryonic feedback may have on the matter power spectrum, currently one of the largest uncertainty in cosmic shear studies at small scales \citep[\eg][]{Chisari2019}.

The halo model power spectrum calculation is known to be inaccurate in detail, particularly in the `transition region' between the two- and one-halo terms \citep{Tinker2005, Valageas2011, Fedeli2014}, which is primarily because linear bias is an overly simplistic treatment of halo--halo correlations \citep{Mead2021b}. 
There are also other errors at larger scales (for example, the lack of perturbation theory) and smaller scales (such as assumed halo sphericity) that arise because of other simplifying assumptions that enter into the standard halo-model calculation. 
To side-step these problems, rather than predicting the raw power, \hmx predicts the ratio of the power spectrum in question to the gravity-only matter--matter spectrum (adopting NFW profiles for all matter) calculated for the same cosmological model. 
The halo-model prediction for this ratio has been shown to be more robust than the absolute power \citep[\eg][]{Cataneo2019} as the underlying cosmology is varied\footnote{\citet{Mead2020} show that this works reasonably well for matter--electron pressure but noted that the transition scale between the two- and one-halo terms occurs at quite different scales depending on the fields being cross-correlated. More work is needed to assess the full impact of this.}. 
Once we have this halo-model ratio, we must multiply by an accurate prescription for the matter--matter power, such as provided by an emulator, or a semi-analytic model, such as \citet{Mead2021-HMCode2020}. 
Here we used the {\sc HMCode-2016} \citep{Mead2016} model for this matter--matter power prediction, since the more accurate {\sc HMCode-2020} \citep{Mead2021-HMCode2020} was not yet available when this analysis was performed.

\subsection{Angular power spectra}
\label{sec:cls}
\begin{figure}
        \begin{center}
                \includegraphics[width=\columnwidth]{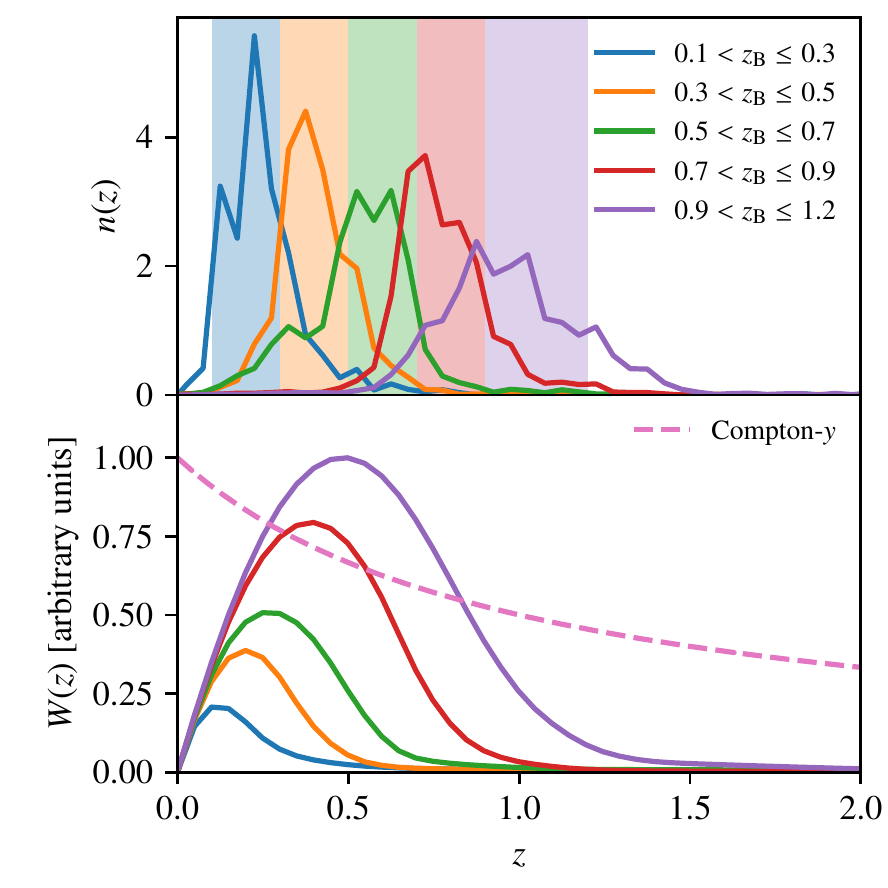}
                \caption{\emph{Top}: Redshift distributions $n_i(z)$ of the five tomographic source samples (solid lines). 
                The shaded bands indicate the selection window on the photometric redshift point estimate $z_\mathrm{B}$. 
                \emph{Bottom}: Line-of-sight window function, $W(z)$, to project the three-dimensional power spectra $P(k)$ (in comoving units) into their angular counterparts, $C_\ell$. 
                The lensing window functions (solid lines, Eq.~\eqref{equ:W-gamma}) are normalised by the maximum of the window function of the last tomographic bin, while the Compton-\y window function (dashed line, Eq.~\eqref{equ:W-y}) is normalised by its maximum value.
                \label{fig:nofz}}
        \end{center}
\end{figure}
Neither the three-dimensional matter--matter nor the matter--electron pressure power spectra are directly observable. 
Instead we observe the angular auto and cross-power spectra of the gravitational lensing shear effect induced by the matter distribution and the tSZ effect caused by the electron pressure. 
The angular power spectrum $C_\ell^{AB}$ at multipole $\ell$ between fields $A$ and $B$ can be related to the three-dimensional power spectrum $P_{AB}(k,z)$ corresponding to these fields through 
\citep[\eg][]{Kaiser1992, LoVerde2008, Kilbinger2017}
\begin{equation}
    \label{equ:cls}
    C_\ell^{AB} = \int_0^{\chi_\mathrm{H}} \frac{\diff\chi}{\chi^2}W^A(\chi)W^B(\chi) P_{AB}\left(\frac{\ell+\frac{1}{2}}{\chi}, z(\chi)\right)\,.
\end{equation}
The integral is taken over the comoving distance $\chi$, from the observer to the distance of the horizon $\chi_\mathrm{H}$. 
Throughout this work we shall assume a flat $\Lambda$ cold dark matter cosmology with a minimal neutrino mass, as there has been no evidence for extensions to this model \citep[\eg][]{Planck2020-Cosmology, Alam2021-eBOSS-cosmology, Troester2021}. 
The shear field derives from the three-dimensional matter overdensity field. 
Hence, for $C_\ell^{\gamma\gamma}$ the power spectrum $P_{AB}$ in Eq.~\eqref{equ:cls} is the matter-overdensity power spectrum $P_\mathrm{mm}$. 
In the case of the angular cross-power spectrum between shear and the tSZ effect, $C_\ell^{\gamma y}$, the relevant three-dimensional power spectrum is the cross-power spectrum of matter overdensity and electron pressure, $P_\mathrm{mP}$. 
Unless noted otherwise, the auto- and cross-correlations involving the shear field shall refer to the $E$-mode of the $E$/$B$-mode decomposition of the spin-2 shear field. 
That is, we shall use $C_\ell^{\gamma\gamma}$ and $C_\ell^{\gamma y}$ as a notational shorthand for $C_\ell^{\gamma_E\gamma_E}$ and $C_\ell^{\gamma_E y}$, respectively. 
The window function $W^{\gamma_i}(\chi)$ of the shear field of the $i$-th source sample is given by
\begin{equation}
    \label{equ:W-gamma}
    W^{\gamma_i}(\chi) = \frac{3}{2}\left(\frac{H_0}{c}\right)^2\Omega_\mathrm{m}\frac{\chi}{a(\chi)} \int_{\chi}^{\chi_\mathrm{H}} \diff \chi'\, n_i(\chi') \frac{\chi'-\chi}{\chi'}\,,
\end{equation}
where $H_0$ denotes the present-day Hubble rate, $c$ the speed of light, $\Omega_\mathrm{m}$ the matter density, $a(\chi)$ the scale factor at comoving distance $\chi$, and $n_i(\chi)\diff\chi = n_i(z)\diff z$ the redshift distribution of source sample $i$. 
The window function $W^y(\chi)$ of the tSZ effect is
\begin{equation}
    \label{equ:W-y}
    W^y(\chi) = \frac{\sigma_\mathrm{T}}{m_\mathrm{e}c^2} \frac{1}{a^2(\chi)}\,,
\end{equation}
where $\sigma_\mathrm{T}$ is the Thompson scattering cross-section and $m_\mathrm{e}$ is the electron mass. 
Here the three-dimensional power spectra are given in comoving units, including the electron pressure. 
That is, the matter--electron pressure cross-power spectrum $P_\mathrm{mP}$ has units of $[(\text{comoving length})^3\times \text{energy}/(\text{comoving length})^3] = [\text{energy}]$. 
The redshift distributions $n_i(z)$ of the source samples, as well as the window functions of Eqs.~\eqref{equ:W-gamma} and \eqref{equ:W-y} are illustrated in Fig.~\ref{fig:nofz}.

In order to compare our model of the angular power spectra in Eq.~\eqref{equ:cls} to the measured power spectra described in Sect.~\ref{sec:summary-stats}, we need to account for the binning in multipoles, residual effects of the mode-mixing matrices, and $E$/$B$-mode mixing.  
To this end we apply the \pCl band-power window function $\mathcal{F}_{b\ell}^{AB}$, which maps the angular power spectrum $C_\ell^{AB}$ to the \pCl band-power power spectrum $C_b^{AB}$ at bin $b$:
\begin{equation}
\label{equ:bandpower-window-functions}
    C_{b}^{AB} = \sum_{\ell} \mathcal{F}_{b\ell}^{AB} C_\ell^{AB}\,.
\end{equation}
The \pCl band-power window function $\mathcal{F}_{b\ell}^{AB}$ can be computed analytically and is given by equation (16) in \citet{Alonso2019}. 
Since our model does not predict any $B$-modes, $E$/$B$-mode mixing does not affect the model for the $E$-modes. 
These window functions are illustrated in Fig. ~\ref{fig:bp-pcl-window-functions} for the case of $E$-mode cosmic shear.

\subsection{Covariance}
\label{sec:covariance}

\begin{figure}
        \begin{center}
                \includegraphics[width=\columnwidth]{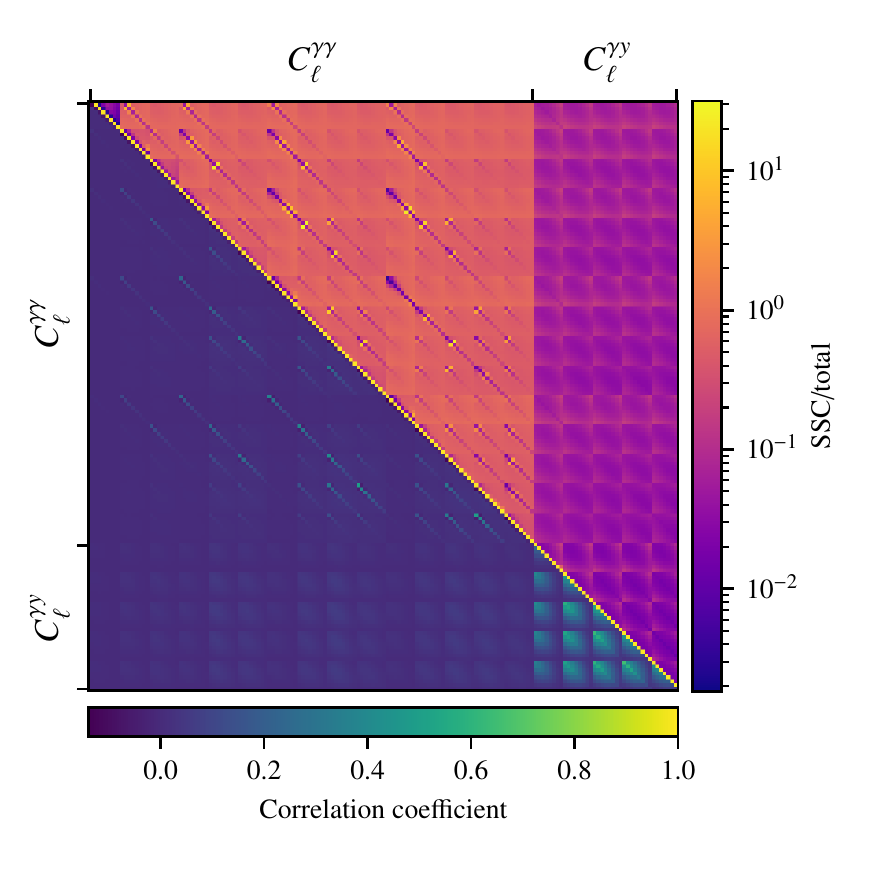}
                \caption{Joint cosmic shear and shear--tSZ data covariance. \emph{Upper right}: Contribution of the super sample covariance (SSC) to the joint covariance. 
                Shown is the ratio between the absolute value of the SSC contribution and the absolute value of the total covariance in Eq.~\eqref{equ:cov-terms}. 
                The ratio can exceed unity, since different contributions to total covariance can cancel one another out to a certain degree. 
                \emph{Lower left}: Correlation coefficients corresponding to the data covariance matrix. 
                \label{fig:corr-coeff}}
        \end{center}
\end{figure}
To estimate the uncertainty associated with our measured data vector, we largely followed the methodology of \citet{Joachimi2021}, with minor modifications to account for the \pCl estimator and allow the computation of the tSZ parts of the covariance matrix. 
In contrast to simulation or resampling-based methods, this analytic approach is free of sampling noise and allows us to cleanly separate the different covariance contributions, as well as match the cosmology to that preferred by the data. 

We decompose the covariance $\mtrx{C} = \Cov{\hat C_{\ell_1}^{ij}, \hat C_{\ell_2}^{kl}}$ of the measured angular (cross-)power spectra $\hat C_\ell^{ij}$ between fields $i$ and $j$ at the effective multipole $\ell$ into
\begin{equation}
    \label{equ:cov-terms}
    \mtrx{C} = \mtrx{C}_\mathrm{G} + \mtrx{C}_\mathrm{SSC} + \mtrx{C}_\mathrm{T} + \mtrx{C}_\mathrm{m}\,,
\end{equation}
where $\mtrx{C}_\mathrm{G}$ is the Gaussian contribution, $\mtrx{C}_\mathrm{SSC}$ is the super-sample covariance (SSC), $\mtrx{C}_\mathrm{T}$ is the non-Gaussian contribution due to the non-zero trispectrum, and $\mtrx{C}_\mathrm{m}$ accounts for the uncertainty in the shear calibration.

We computed the Gaussian covariance term $\mtrx{C}_\mathrm{G}$ using \software{NaMaster}, accounting for the effect of partially overlapping and weighted sky observations \citep{Garcia-Garcia2019}. 
While this can be done exactly, the calculation scales as $O(\ell_\mathrm{max}^6)$ for the sample variance term. 
Instead, we used the improved narrow kernel approximation (NKA) introduced in \citet{Nicola2021} for the cosmic shear sample variance and the parts of the covariance that involve tSZ cross-correlations.
For cosmic shear, the noise terms can be computed exactly with little extra effort by calculating the different terms in Eq.~(2.29) of \citet{Nicola2021} separately. 
We found that this exact calculation reduces the Gaussian cosmic shear covariance by up to 20\,\% compared to the case where the NKA is used for both sample variance and noise terms. 
Some of the Compton-\y maps have no readily available noise maps necessary for the exact noise computation, and we thus used the NKA for all terms of covariance involving the tSZ effect. 
As the signal-to-noise ratio of the tSZ cross-correlation is lower than that of the cosmic shear measurement, this is of little concern, however. 
The sample variance of the cosmic shear and shear--tSZ terms in the Gaussian covariance were computed using the model for the angular power spectra of Sect.~\ref{sec:cls} at the best-fit cosmology of \citet{Heymans2021}. 
By contrast, the tSZ auto-power spectrum term is based on a smoothed version of the measured auto-power spectrum of the \y-maps. 
This choice is motivated by the fact that \hmx was not calibrated for accurate predictions of the pressure auto-power spectrum. 
Since the overlap between the KiDS-1000 footprint and the Compton-\y maps constitutes a small fraction of the full \y-maps, the measured tSZ auto-spectra are largely independent of the shear spectra.

\citet{Osato2021} find that the SSC term can be the dominant contribution to the data covariance in analyses of the tSZ auto-power spectrum. 
In cosmic shear analyses, the SSC term is the dominant non-Gaussian contribution \citep[\eg][]{Barreira2018}. 
We thus accounted for SSC in the full cosmic shear, shear--tSZ, and joint covariance. 
The SSC $\mtrx{C}_\mathrm{SSC}$ and trispectrum terms $\mtrx{C}_\mathrm{T}$ were computed using the halo model formalism as implemented in \software{CCL}\footnote{\url{https://github.com/LSSTDESC/CCL}} \citep{Chisari2019-CCL}. 
The SSC terms of the cosmic shear and shear--tSZ covariances used the power and cross-power spectra of the respective survey masks, while the SSC term of the cross-covariance between the two data vectors used the cross-power spectrum of shear and tSZ masks. 
The only terms in the full covariance Eq.~\eqref{equ:cov-terms} that did not depend on the shape of the survey footprints are the trispectrum terms, which used the effective survey areas and overlap-areas instead. 
The halo matter density and pressure profiles were computed with the \hmx prescription, using \software{pyhmcode}\footnote{\url{https://github.com/tilmantroester/pyhmcode}}, such that the modelling of the signal and the covariance is consistent.
The version of \software{CCL} used in this work only includes the one-halo contribution to the trispectrum term. 
This is not an issue, however, since for cosmic shear the trispectrum term is subdominant everywhere \citep{Joachimi2021} and for the tSZ effect it is dominated by the one-halo term \citep[][]{Cooray2001b}.  
The SSC term was found to be sub-dominant for the shear--tSZ cross-correlation but can reach up to $\sim 20\,\%$ of the total covariance for off-diagonal terms. 
This is illustrated in Fig.~\ref{fig:corr-coeff}, showing the relative contribution of the SSC to the total covariance. 
We found that not accounting for the SSC term led to an underestimate of the inferred parameter uncertainties of about 10\,\%. 

To account for the uncertainty of the multiplicative shear bias $\mtrx{C}_\mathrm{m}$, we followed \citet{Joachimi2021}, that is, we assumed that the multiplicative bias corrections are fully correlated between the tomographic bins. 
\citet{Asgari2021-CS} show that this approach gives very similar constraints to allowing the multiplicative biases to vary, while not requiring sampling over an extra five parameters. 

Another approach to the estimation of the data covariance is through large ensembles of simulated mock data. 
This approach has some considerable downsides for the present analysis, however. 
Obtaining mock data that accurately models both the tSZ effect, as well as weak lensing, over large sky areas and with a sufficient number of realisations to reduce the sampling noise, is a challenging proposal. 
While \citet{Troester2019} demonstrate a deep-learning method to augment weak-lensing-optimised \nbody simulations with gas physics that accurately predict the tSZ effect, this approach is not easily amenable to different cosmological parameters. 
Furthermore, accounting for the SSC is difficult, unless one employs costly `separate universe' simulations \citep{Li2014}. 
Finally, the covariance is dominated by the Gaussian term \citep[][]{Joachimi2021}, which can be accurately computed analytically, even for large survey areas.

Our data covariance model was checked against jackknife estimates based on various patch sizes and jackknife estimators. 
We found agreement at the 10\,\% level on average but with large scatter between different jackknife estimates. 
The jackknife approaches tended to slightly overestimates (underestimates) the covariance at small (large) scales. 
However, since the jackknife patches are heteroskedastic due to inhomogeneous masks and weight maps, as well as breaking the assumption of independence between the jackknife patches at large scales, these jackknife estimates need to be interpreted with caution. 

\subsection{Systematics}
\label{sec:systematics}

\subsubsection{Photometric redshift uncertainties}
\label{sec:photoz}
The redshift distributions of the \KiDS source samples have been estimated and validated to a high degree of accuracy \citep{Hildebrandt2021-NZ}. 
To account for the residual uncertainty in the redshift distributions of the source samples, we followed \citet{Asgari2021-CS} and allowed the redshift distribution for each tomographic bin to shift. 
We imposed a multivariate Gaussian prior on these shifts, where the mean and covariance were derived from the calibration of the redshift estimation method to realistic mock data. 
We refer the interested reader to \citet{Wright2020}, \citet{Hildebrandt2021-NZ}, and \citet{Asgari2021-CS} for further details on the estimation of the redshift distributions and their uncertainties.

\begin{figure}
        \begin{center}
                \includegraphics[width=\columnwidth]{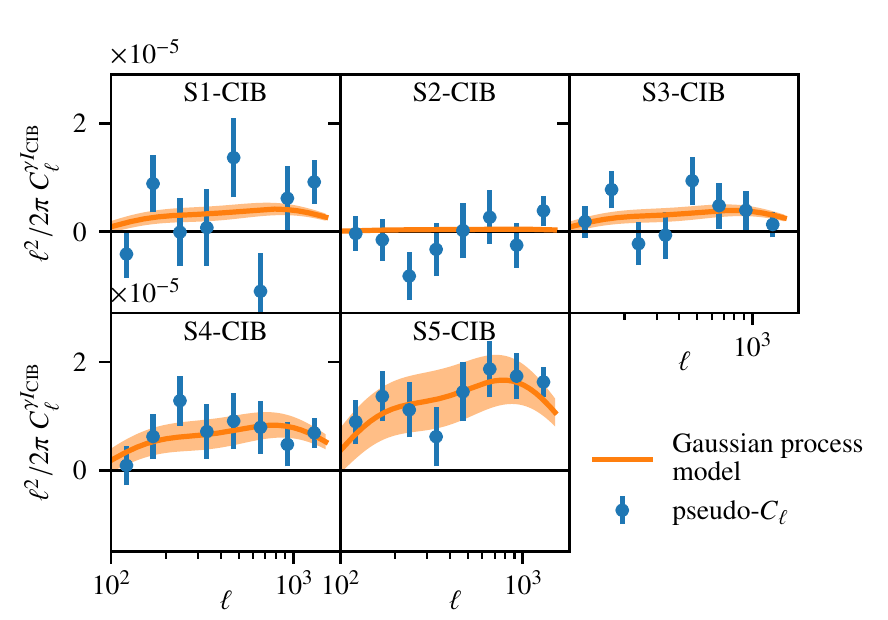}
                \caption{Measured angular cross-power spectra between KiDS-1000 shear and the $545\,\mathrm{GHz}$ CIB map (blue points). 
                The best-fit Gaussian process (GP) model is shown as the orange line, with the orange shaded band indicating the 68th percentiles of the GP model uncertainty. 
                \label{fig:CIB-GP}}
        \end{center}
\end{figure}

\subsubsection{Intrinsic alignments}
Since the tSZ effect is correlated with large-scale structure, source galaxies aligning with the same large-scale structure can mimic a gravitational lensing signal, and thus the correlation between the tSZ effect and these IA effects could bias our shear--tSZ cross-correlation analysis. 
We accounted for this IA effect using the same model as in the KiDS-1000 analyses \citep[see][]{Joachimi2021, Asgari2021-CS, Heymans2021}. 
Specifically, we employed the non-linear alignment \citep[NLA;][]{Hirata2004, Bridle2007} model and allowed the amplitude parameter $A_\mathrm{IA}$ to vary, while keeping the redshift-evolution parameter $\eta_\mathrm{IA}$ fixed. 
\citet{Asgari2021-CS} show that letting $\eta_\mathrm{IA}$ vary serves to slightly loosen the constraints on $S_8$ but does not affect its central value. 
For a detailed discussion of the choice of IA model for the KiDS-1000 data, we refer the interested reader to appendix~D of \citet{Heymans2021}.
We found that IA has a small but significant impact on the shear--tSZ cross-correlation signal, in agreement with findings of \citet{Gatti2021} and \citet{Pandey2021}. 
Its relative contribution to the total signal, as well as its fraction of the measurement error, is comparable to the IA contribution in the case of cosmic shear.

\subsubsection{Cosmic infrared background}
\label{sec:CIB}
The tSZ maps are contaminated by CIB \citep{Hurier2015, Planck2015-tSZ, Planck2015-tSZxCIB}.  %
To assess and correct for this contamination in our shear--tSZ cross-correlation, we considered two complementary approaches. 
The first approach models the effect of the CIB contamination based on the observed correlation between the shear field and maps of the CIB, while the second approach uses Compton-\y maps where the CIB has been subtracted during the reconstruction process.

For the first, model-based approach, we followed the simple but commonly used method of \citet{Hill2014} and \citet{Vikram2017}, where the contamination of the observed Compton-\y map $y_\mathrm{obs}$ is modelled as
\begin{equation}
    \label{equ:y_cont}
    y_\mathrm{obs} = y_\mathrm{true} + \alphaCIB I_\mathrm{CIB} + N\,,
\end{equation}
where $y_\mathrm{true}$ is the map of the uncontaminated tSZ effect, $I_\mathrm{CIB}$ is a map tracing the CIB, \alphaCIB is a scalar parameter, and $N$ denotes the collection of noise and other contaminants that are not expected to correlate with the shear field, unlike the other two terms. 

This model is quite simplistic; the use of a single frequency CIB map assumes that the CIB is not varying with frequency, and the single, scalar parameter \alphaCIB assumes that the contamination is not varying spatially.
Neither of these assumptions strictly hold true: the CIB is the result of complex astrophysical processes, whose detailed frequency dependence is not constant, and the reconstruction algorithms used for the \y-maps in this work employ scale-dependent filters, resulting in scale-dependent ILC coefficients, and would thus require a scale-dependent \alphaCIB as well. 

In practice, these assumptions appear to be sufficient, however, at least for the applications considered in the present work, since we found CIB contamination to be negligible (see Sect.~\ref{sec:CIB-results}). 
Compton-\y maps using scale-independent ILC weights \citep[e.g.][]{Van-Waerbeke2014, Hill2014} are consistent with those using \milca and \nilc reconstructions used here and the CIB is strongly correlated between different frequency bands \citep[\eg][]{Planck2013-CIB, Lenz2019, Stein2020}.
Nonetheless, future work will have to revisit these approximations. 
Potential improvements include the use of a multi-band analysis, such as in \citet{Chiang2020}, where the signal in each frequency band is modelled separately, or the use of other CIB maps \citep[e.g.][]{Lenz2019}.

Under the model of Eq.~\eqref{equ:y_cont}, the cross-power spectrum between the shear field and the observed Compton-\y map is then given by
\begin{equation}
    C_\ell^{\gamma y_\mathrm{obs}} = C_\ell^{\gamma y_\mathrm{true}} + \alphaCIB C_\ell^{\gamma I_\mathrm{CIB}}\,,
\end{equation}
where $C_\ell^{\gamma y_\mathrm{true}}$ corresponds to the shear--tSZ cross-correlation model in the absence of CIB contamination and $C_\ell^{\gamma I_\mathrm{CIB}}$ is a model for the shear-CIB cross-correlation. 
The scaling parameter \alphaCIB can be found by comparing the measured cross-correlation between the Compton-\y map and the CIB map with their auto-correlations.
Here we used the $545\,\mathrm{GHz}$ CIB map from \citet{Planck2015-CIB} as a proxy for $I_\mathrm{CIB}$ in Eq.~\eqref{equ:y_cont} together with {$\alphaCIB = (2.3\pm6.6)\times 10^{-7}\,\mathrm{MJy}^{-1}\mathrm{sr}$}, as derived in \citet{Alonso2018}.

To model $C_\ell^{\gamma I_\mathrm{CIB}}$ we chose a phenomenological approach, since modelling the CIB within the \hmx framework is not possible at this point. 
Instead, we measured the tomographic cross-power spectra between the KiDS-1000 source samples and the $545\,\mathrm{GHz}$ CIB map (smoothed to a beam size of $10'$ and masked with the \y-map mask) and then fitted a Gaussian process (GP) to these observed cross-power spectra.
The shear--CIB cross-spectra, together with the GP model and its uncertainty are shown in Fig.~\ref{fig:CIB-GP}.
The Gaussian data covariance was computed in an iterative process. 
First, the covariance was computed by \software{NaMaster} using the cosmic shear model of Sect.~\ref{sec:cls} and the measured and smoothed auto-power spectrum of the CIB map, with zero cross-correlation. 
After fitting the GP model using this initial covariance, the covariance was recomputed, now including the GP model for the cross-correlation term. 
Finally, the GP model was refitted using the new covariance. 
The GP model was implemented using \software{gpytorch}\footnote{\url{https://gpytorch.ai/}} \citep{Gardner2018-gpytorch}. 
The model is a multi-task GP \citep{Bonilla2008}, with a fixed data covariance. 
The fitted parameters consist of the length-scale of the squared exponential kernel, which accounts for the correlation between $\ell$-modes, and the task covariance, which accounts for the correlation between the tomographic bins and is parameterised as the sum of a rank-one matrix and a diagonal component, for a total of 11 free parameters.
Being a multi-task GP, the model thus accounts for the correlation between the redshift bins, as well as between $\ell$-bins. 
Using a fixed data covariance ensures that the GP regression does not fit to spurious or excess (co)-variances. 
This formalism can be extended to account for the correlation between the shear--tSZ and shear--CIB cross-correlations, such that the GP model is conditioned on the shear--tSZ model. 
Due to the small effect of the CIB on the present analysis, we leave this sophistication to future work.

To account for residual biases in our treatment of the CIB contamination, we marginalised over the $\alphaCIB$ parameter in our fiducial analysis, using a Gaussian prior based on the value derived in \citet{Alonso2018} but with a variance that is four times larger. 
Alternatively, being a linear parameter, the marginalisation over $\alphaCIB$ could have been accounted for analytically by an appropriate term in the data covariance \citep[\eg][]{Bridle2002}, analogous to the treatment of the multiplicative shear bias. 

\begin{figure}
        \begin{center}
                \includegraphics[width=\columnwidth]{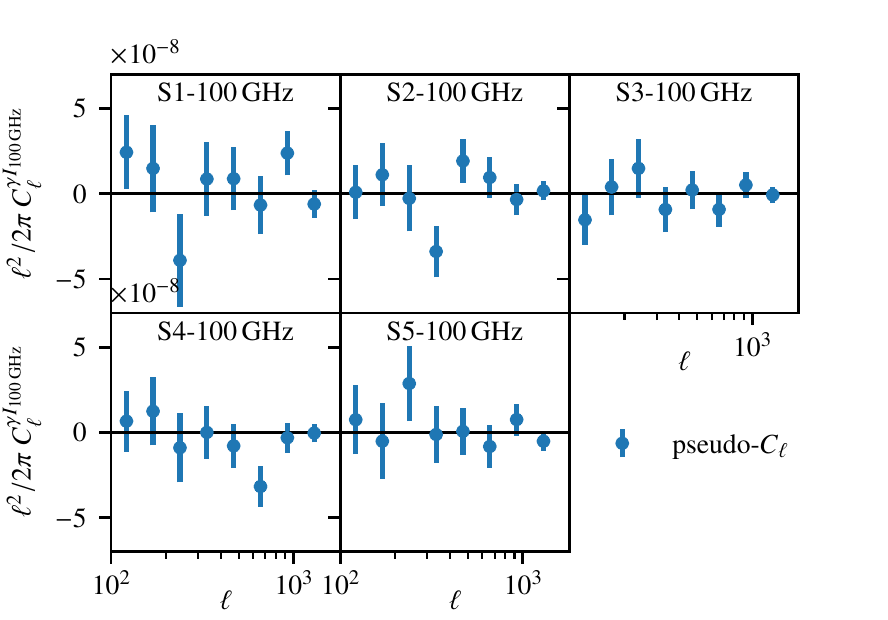}
                \caption{Measured angular cross-power spectra between KiDS-1000 shear and the \Planck $100\,\mathrm{GHz}$ intensity map.
                \label{fig:100GHz-measurements}}
        \end{center}
\end{figure}

In our second approach to assess the sensitivity of our cross-correlation measurements to CIB contamination, we computed the shear--tSZ cross-correlation using CIB-subtracted \y-maps from \citet{Yan2019}. 
These maps subtract the CIB contributions in the $353\,\mathrm{GHz}$, $545\,\mathrm{GHz}$, and $845\,\mathrm{GHz}$ bands, based on the CIB maps constructed in \citet{Planck2015-CIB}. 
The CIB contributions at lower frequencies are extrapolated from these maps, using a grey body spectrum with the best-fit parameters from \citet{Planck2013-CIB}. 

A different approach was used in the construction of the ACT Compton-\y map \citep{Madhavacheril2020} and the \y-maps of the recent Dark Energy Survey Year 3 (DES Y3) cross-correlation analyses \citep{Gatti2021, Pandey2021}. 
During the ILC procedure, rather than projecting out the primary CMB, they deproject a grey body spectrum. 
The grey body spectrum is parameterised by the dust temperature $T_\mathrm{d}$ and the emissivity index $\beta$ of the grey body spectrum, which modifies the slope of the black-body spectrum at low frequencies.
\citet{Gatti2021} and \citet{Pandey2021} varied the emissivity index $\beta$ between 1 and 1.4. 
This is significantly lower than the $\beta=1.75\pm 0.06$ derived from the CIB power spectrum in \citet{Planck2013-CIB} and which we used in our CIB-subtracted \y-maps.
Other works, such as the WebSky mocks \citep{Stein2020} use the model parameters derived in \citet{Viero2013} based on \textit{Herschel} CIB data, which prefer $\beta=1.6\pm 0.1$. 
The effective dust temperature is $T_\mathrm{d} = 24\,\mathrm{K}$ in \citet{Gatti2021}, while \citet{Yan2019} use an effective dust temperature at $z=1.2$ of $T_\mathrm{d} = 32.4\,\mathrm{K}$. 
However, at \Planck frequencies the effect of the dust temperature and its redshift evolution is subdominant to the large changes in the emissivity index considered here. 
To check our CIB-subtracted Compton-\y map against changes in $\beta$, we also generated a map that extrapolates the CIB to the lower frequency bands assuming a value of $\beta=1.2$. 
In Sect.~\ref{sec:CIB-results}, we show that our results are robust with respect to all of these choices.

\subsubsection{Radio sources}
Unresolved radio sources are another potential contamination in the Compton-\y map. 
\citet{Shirasaki2019} suggest that unresolved radio sources can bias shear--tSZ cross-correlations by up to 30\,\%. 
Their study however differs in key assumptions from ours. 
To calculate the expected contamination from unmasked radio sources, they assumed a flux cut of $400\,\mathrm{mJy}$. 
In our analysis, we used the point-source and 40\,\% Galactic mask provided in \citet{Planck2015-tSZ}. 
The mask covers sources with fluxes $> 300\,\mathrm{mJy}$ of the Second \Planck Catalogue of Compact Sources \citep[PCCS2;][]{Planck-Collaboration2016-PCCS2}, with none of the unmasked sources lying within the KiDS-1000 footprint. 
The flux cut assumed in \citet{Shirasaki2019} therefore overestimates the number and flux from unmasked sources. 
Considering the sizeable bias estimated in \citet{Shirasaki2019}, even a lower flux cut might still give rise to a significant amount of radio contamination in our Compton-\y maps. 
We therefore followed a similar approach to our treatment of the CIB contamination, namely estimate the effect on the shear--tSZ cross-correlation based on the cross-correlation of the shear field with the contaminant. 
To this end, we measured the angular cross-power spectrum between the KiDS-1000 source samples and the \Planck $100\,\mathrm{GHz}$ map as a tracer of radio emissions. 
The cross-correlation with Gaussian errors is shown in Fig.~\ref{fig:100GHz-measurements}. 
Unlike the cross-correlation with the CIB, we detected no significant correlation between the shear field and the $100\,\mathrm{GHz}$ map. 
We hence concluded that our estimates of the shear--tSZ cross-correlation are not significantly affected by contamination of our \y-maps due to unresolved radio sources. 
This finding is also in agreement with \citet{Planck2015-tSZ}, who find that radio sources are a subdominant contribution to the Compton-\y auto-power spectrum, as well as with \citet{Gatti2021}, who find that masking radio sources does not affect their shear--tSZ cross-correlation.

\begin{table*}
\caption{Sampled parameters and priors.}              
\label{tab:priors}      
\centering                                      
\begin{tabular}{lll}          
\toprule
Parameter & Symbol & Prior \\    
\midrule                                   
Dimensionless Hubble constant & $h$ & $U(0.64,\,0.82)$ \\
Physical baryon density & $\omega_{\rm b}$ & $U(0.019,\,0.026)$ \\
CDM density & $\omega_{\rm c}$ & $U(0.051,\,0.255)$ \\
Density fluctuation amplitude & $S_8$ & $U(0.1,\,1.3)$ \\
Scalar spectral index & $n_{\rm s}$ & $U(0.84,\,1.1)$ \\
\midrule
Intrinsic alignment amplitude & $A_{\rm IA}$ & $U(-6,\,6)$ \\
CIB contamination amplitude & $\alpha_\mathrm{CIB}$ [$10^7$ MJy$^{-1}$ sr] & $\mathcal{N}\left(2.3, (13.2)^2\right)$ \\
AGN feedback strength & $\logtheat$ & $U(7.1,\, 8.5)$ \\
Redshift offsets (5) & $\vec{\delta_z}$ & ${\cal N}(\vec{\mu}_{\delta_z}, C_{\delta_z})$ \\
\bottomrule
\end{tabular}
\tablefoot{The first section lists the primary cosmological parameters, and the second section lists the astrophysical and observational nuisance parameters for intrinsic galaxy alignments, CIB contamination, baryon feedback, and uncertainties in the redshift calibration. 
The five redshift offset parameters are drawn from a multivariate Gaussian prior with mean $\vec{\mu}_{\delta_z}$ and covariance $C_{\delta_z}$ (see Sect.~\ref{sec:photoz} for details). 
}
\end{table*}

\subsection{Inference pipeline}
Our inference pipeline is based on a modified version\footnote{\url{https://github.com/KiDS-WL/kcap}} of \software{CosmoSIS}\footnote{\url{https://bitbucket.org/joezuntz/cosmosis}} \citep{Zuntz2015}. 
The linear matter power spectrum is calculated using \software{CAMB}\footnote{\url{https://github.com/cmbant/CAMB}} \citep{Lewis2000}. 
The non-linear matter and electron pressure power spectra are computed using \hmx, after which the modelling proceeds as described in Sects.~\ref{sec:cls} and \ref{sec:systematics}. 
The computational cost of the model prediction is dominated by the calculation of the non-linear power spectra, specifically the calculation of the halo profiles, since the matter and electron-pressure profiles have no analytic Fourier transforms, necessitating numerical transforms. 
To mitigate this, we evaluated the three-dimensional power spectra at eight redshifts between $z=0$ and $z=1.5$, which reduces the evaluation time of a single likelihood to $\sim 30\,\mathrm{s}$.  
The smoothness of the power spectra and kernels in the integral of Eq.~\eqref{equ:cls} limits the numerical error incurred by this coarse redshift sampling. 
We found that this redshift sampling is accurate to 1\,\% of the measurement uncertainty for the shear--tSZ cross-correlation and to 0.5\,\% for cosmic shear. 
We assumed a Gaussian likelihood, given by the model described above; a data vector consisting of either the cosmic shear or shear--tSZ data, as described in Sect.~\ref{sec:summary-stats}, or the concatenation of the two in the case of the joint analysis of cosmic shear and shear--tSZ; and the data covariance described in Sect.~\ref{sec:covariance}. 

In our fiducial setup, we sampled 13 parameters, listed in Table~\ref{tab:priors}, using the nested sampling implementation of \software{MultiNest} \citep{Feroz2008,Feroz2009,Feroz2013}, using 500 or 1000 live points, and an efficiency parameter of 0.3. 
The maximum of the posterior was found by running multiple Nelder-Mead \citep{Nelder1965} optimisations, starting from samples with the highest posterior values.

\begin{figure*}
        \begin{center}
                \includegraphics[width=\textwidth]{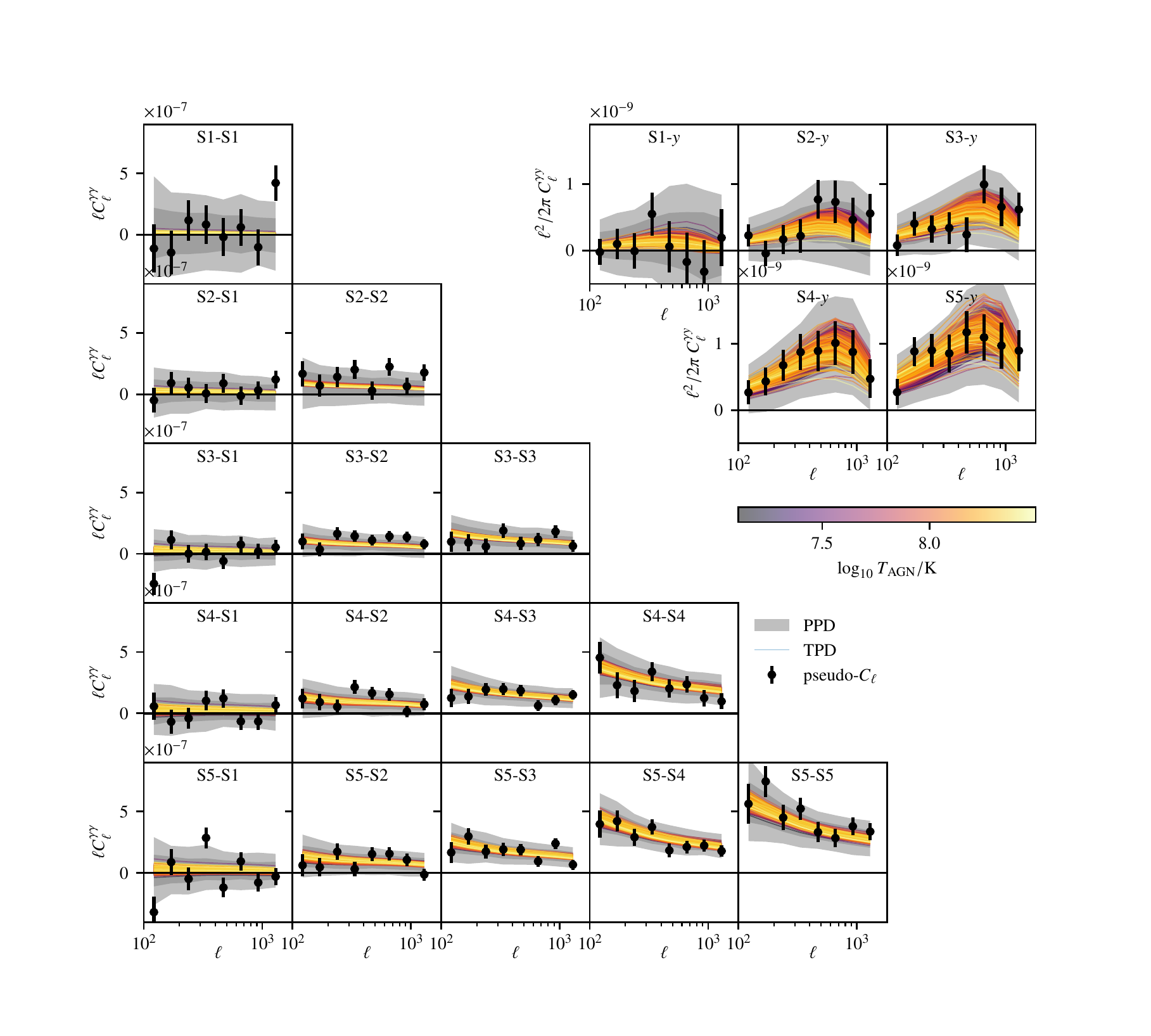}
                \caption{Measured angular power spectra of KiDS-1000 cosmic shear (bottom left) and the angular cross-power spectra between KiDS-1000 shear and the tSZ effect (top right). 
                Shown here is the cross-correlation using the {\milca} \Planck Compton-\y map. 
                The grey bands indicate the 68th and 95th percentile of the posterior predictive distribution (PPD) of the joint-analysis posterior. 
                Coloured lines are samples of the translated posterior distribution (TPD), with the colour corresponding to the baryon feedback strength. 
                Since the TPD is derived from the full joint-posterior, the baryon feedback strength only accounts for part of the variance of the shown TPD samples.
                \label{fig:best-fit-model}}
        \end{center}
\end{figure*}

\section{Results}
\label{sec:results}

\begin{figure}
        \begin{center}
                \includegraphics[width=\columnwidth]{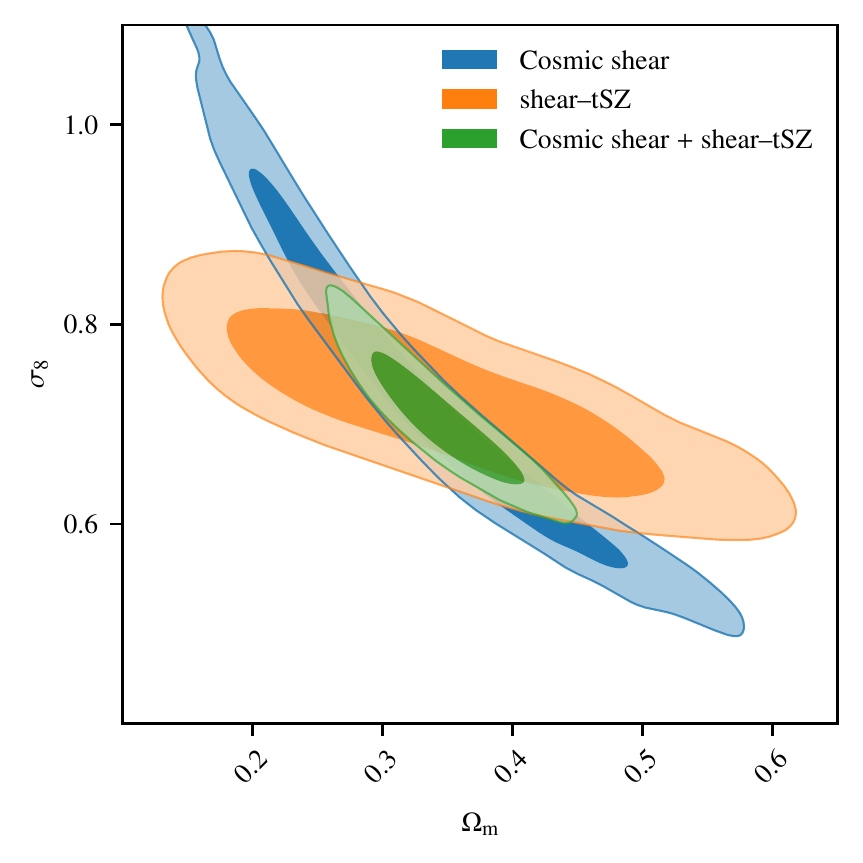}
                \caption{Constraints from KiDS-1000 cosmic shear (blue), the cross-correlation of shear and the tSZ effect from \Planck (orange), and their joint analysis (green) on \om and $\sigma_8$.
                \label{fig:constraints-banana}}
        \end{center}
\end{figure}

\begin{figure}
        \begin{center}
                \includegraphics[width=\columnwidth]{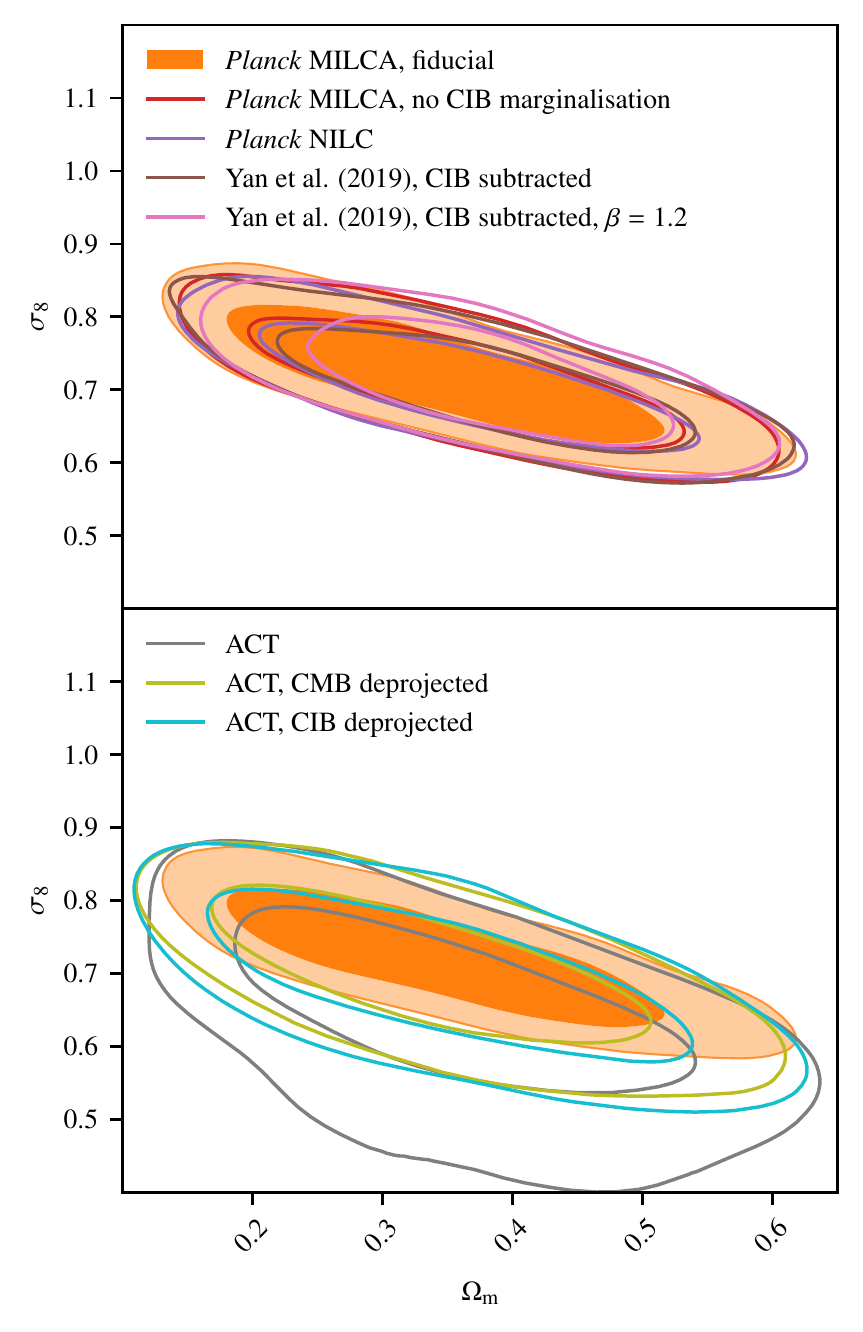}
                \caption{Effect of different Compton-\y map reconstruction methods on the shear--tSZ cross-correlation constraints on \om and $\sigma_8$. 
                \emph{Top}: Compton-\y maps derived from \Planck data. 
                \emph{Bottom}: Compton-\y maps from ACT. 
                \label{fig:constraints-CIB-banana}}
        \end{center}
\end{figure}

\begin{figure}
        \begin{center}
                \includegraphics[width=\columnwidth]{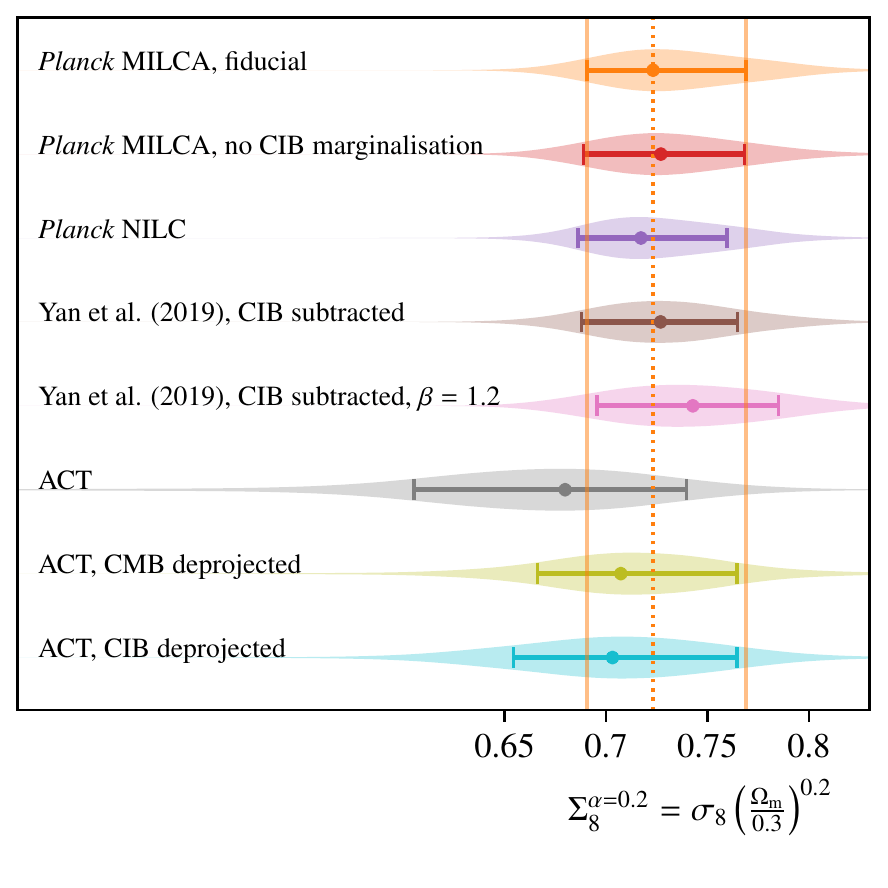}
                \caption{Marginal posterior densities of $\Sigmaalpha=\sigma_8\left(\om/0.3\right)^{0.2}$ from the cross-correlation of KiDS and different Compton-\y maps. 
                The orange band and dashed line indicate the marginal 68th percentile CI and mode of the marginal posterior of the fiducial setup.
                \label{fig:constraints-Sigmaalpha}}
        \end{center}
\end{figure}

Figure~\ref{fig:best-fit-model} presents the measurements of our fiducial data vector, consisting of the \pCl estimates of the KiDS-1000 cosmic shear angular power spectra and the shear--tSZ angular cross-power spectra, using the {\Planck} {\milca} Compton-\y map. 
Overplotted are the 68 and 95 percentiles of the posterior predictive distribution (PPD) and samples of the translated posterior distribution \citep[TPD;][]{Kohlinger2019} for the joint data vector, coloured by their value of \logtheatinline, which characterises the strength of AGN feedback in our model. 
The PPD describes fictional observations, conditioned on the data and model, and is a useful diagnostic to assess consistency within and between data vectors \citep[\eg][]{Gelman2013,Kohlinger2019, Doux2021}. 
The TPD translates the model parameter posterior into data-space, thus capturing the distribution of model predictions, conditioned on the data. 

The cosmic shear signal is detected at \SNEE signficance, while the shear--tSZ cross-correlation is detected at \SNTEMILCA significance. 
The significance was estimated as $\sqrt{\vec d^\mathrm{T} \mtrx{C}^{-1}\vec d}$, where $\vec d$ is the \pCl data vector and $\mtrx{C}$ is the data covariance, given by Eq.~\eqref{equ:cov-terms}. 
The $B$-modes of our measurements are consistent with zero, with details discussed in Appendix~\ref{sec:b-modes}.

Parameter constraints on \om and $\sigma_8$ derived from the cosmic shear, shear--tSZ, and joint data are shown in Fig.~\ref{fig:constraints-banana}. 
Our cosmic shear constraints exhibit the usual `cosmic banana' degeneracy and are in good agreement with the KiDS-1000 band-power analysis of \citet[][cf. Appendix~\ref{app:bp-vs-pcl}]{Asgari2021-CS}. 
The cosmic shear signal scales approximately as $\propto \sigma_8^2\om$ and is thus usually summarised with the parameter $S_8 = \sigma_8\sqrt{\om/0.3}$. 
Depending on the specific estimator, the exact scaling can deviate slightly from $S_8$, such that the parameter $\Sigma_8^\alpha = \sigma_8\left(\om/0.3\right)^{\alpha}$, with an appropriately chosen $\alpha$, is a better description of the direction orthogonal to the degeneracy. 
For example, the $\sigma_8$--\om degeneracy for the band-power estimator in \citet{Asgari2021-CS}, as well as the \pCl estimator in this work, is better described by $\Sigma_8^{0.58}$ than $S_8$. 

The shear--tSZ cross-correlation is more sensitive to $\sigma_8$ \citep[\eg][]{Refregier2002, Komatsu2002, Mead2020} than cosmic shear, resulting in a degeneracy direction that is shallower in $\sigma_8$. 
We find that the cross-correlation signal scales approximately as $\propto \sigma_8^5\om$.
We use the parameter $\Sigma_8^{0.2}$ when comparing different shear--tSZ cross-correlation constraints in the remainder of this analysis. 
In our fiducial analysis, which uses the {\Planck} \milca Compton-\y map, we infer
\begin{equation}
    \Sigmaalpha = \SigmaalphaXcorr\,.
\end{equation}

The different degeneracy directions of the cosmic shear and shear--tSZ cross-correlations allow the joint analysis of the two to break these degeneracies and provide tight cosmological constraints. 
We find
\begin{splitequation}
    \om &= \omJoint\, \\
    \sigma_8 &= \sigmaeightJoint\, \\
    S_8 &= \SeightJoint\,
\end{splitequation}
for the maximum of the marginal posterior and marginal highest posterior density credible intervals (CIs).  
The $S_8$ constraints are improved by a factor of \SeightJointImprovement over cosmic shear alone and are in excellent agreement with the KiDS-1000 {$3\times2$pt} analysis of \citet{Heymans2021}. 
This improvement is driven by the breaking of the residual degeneracy between $S_8$ and \om, since $S_8$ does not optimally capture the direction orthogonal to the $\sigma_8$--\om degeneracy for the estimator considered here. 
Using instead $\Sigma_8^{0.58}$, as in \citet{Asgari2021-CS}, which is better constrained than $S_8$ by the \pCl estimator, we find a \SigmaalphaJointImprovement improvement of the constraining power of the shear--tSZ + cosmic shear joint analysis over cosmic shear alone. 

The joint analysis also strongly tightens the constraints on \om and $\sigma_8$. 
This is to be expected, however, since the cosmic shear constraints on these parameters are prior-dominated, and any breaking of the cosmic banana will yield comparatively tight constraints compared to cosmic shear alone \citep[see \eg][for examples of this effect in the joint analyses of KiDS-1000 cosmic shear with CMB lensing or super novae]{Troester2021}. 

The shear--tSZ cross-correlation prefers lower values of $\sigma_8$ than those inferred from the CMB by \Planck; the discrepancy between our inferred value of $\Sigmaalpha$ and that of \citeauthor{Planck2020-Cosmology} (\citeyear{Planck2020-Cosmology}; using the TTTEEE+lowE likelihood) is $2.2\,\sigma$. 
This trend is also seen in the recent shear--tSZ cross-correlation analysis with DES Y3 data \citep{Gatti2021}, although they impose \Planck priors on $\sigma_8$ and \om, making a direct comparison with the results in this work difficult. 
The constraints on $S_8$ and $\Sigma_8^{0.58}$ inferred from the joint analysis are in $3.4\,\sigma$ and $2.9\,\sigma$ tension with those of \Planck, similar to previous joint analyses of KiDS-1000 with other datasets \citep{Heymans2021, Troester2021}, as well as KiDS-1000 cosmic shear by itself \citep{Asgari2021-CS}. 
The full parameter-space constraints are shown in Fig.~\ref{fig:constraints-all}. 

We note here that the considerable challenges of the accurate and consistent modelling of the cross-correlation signal warrant revisiting these results in the future, when the modelling has matured and more data have become available. 
In particular, accounting for the dependences of, and interactions between, cosmology, baryon feedback, and IA requires both models that consistently include these effects and simulations that sufficiently cover this parameter space to validate said models. 
These points are further discussed in Sect.~\ref{sec:baryons-results}. 

\subsection{Goodness of fit}
We find that the data are well described by our model. 
The $\chi^2$ of the \pCl cosmic shear data with respect to the model of the maximum a posteriori (MAP) estimate is \chisqCS, with \ndofCS effective degrees of freedom. 
The effective model dimensionality was estimated following \citet{Raveri2019}, and accounts for the fact that many of the varied parameters in the model are prior-dominated and as such do not contribute extra degrees of freedom to the model. 
Our fiducial shear--tSZ cross-correlation has a $\chi^2$ of \chisqXcorr, with \ndofXcorr effective degrees of freedom, while the joint analysis has a $\chi^2$ of \chisqJoint, with \ndofJoint effective degrees of freedom. 
The goodness-of-fit values of our measurements are summarised in Table~\ref{tab:goodness-of-fit}.
 
\subsection{Alternative Compton-\y maps and effect of the CIB}
\label{sec:CIB-results}
To assess the degree to which \y-map reconstruction and CIB contamination treatment affect our results, we derived constraints using a range of Compton-\y maps and compared the resulting \om-$\sigma_8$ and \Sigmaalpha constraints to those derived based on our fiducial setup, consisting of the {\Planck} \milca \y-map and CIB marginalisation using the GP model described in Sect.~\ref{sec:CIB}. 
The \om-$\sigma_8$ constraints based on the other Compton-\y maps are shown in Fig.~\ref{fig:constraints-CIB-banana}. 
The constraints on \Sigmaalpha are compared in Fig.~\ref{fig:constraints-Sigmaalpha}. 

The \Planck-derived maps yield tighter constraints than those from ACT, owing to the larger overlap between the KiDS-1000 and \Planck data. 
Among the ACT maps, we find that the \y-map where no other component has been deprojected leads to weaker constraints, with a tail to low values of \Sigmaalpha. 
We explain this effect with a slightly larger data covariance due to a higher auto-power spectrum, as well as anomalously low data points in the shear--tSZ cross-correlation of the Compton-$y$ map with no deprojected component compared to the maps with either the CMB or CIB projected out. 
We discuss these points in more detail in Appendix~\ref{app:ACT-maps}. 
Both the CMB-deprojected and CIB-deprojected \y-maps give very similar constraints, with no systematic shift between them. 
If the cross-correlation was significantly affected by CIB contamination, we would expect to find different constraints in these two cases. 

We find good agreement between all maps and no evidence that the CIB causes significant biases in our constraints. 
This is in contrast to the finding in \citet{Gatti2021}, who find that deprojecting a grey body spectrum during the construction of their \Planck-derived Compton-\y map leads to significantly different cross-correlation signals compared to the case where no such deprojection was applied. 
The origin of this discrepancy is currently unknown, however, and will require a detailed comparison of the different Compton-\y maps. 
We leave this for future work. 

The mask of our Compton-\y maps is potentially correlated with the lensing signal, as the point source mask includes some clusters. 
Since these clusters are sources of both tSZ and weak lensing signals, this could in principle cause the estimated signal to be biased low. 
As the number of masked point sources in the KiDS-1000 footprint is low, this effect is likely not relevant in our case.

\subsection{Baryon feedback}
\label{sec:baryons-results}
\begin{figure}
        \begin{center}
                \includegraphics[width=\columnwidth]{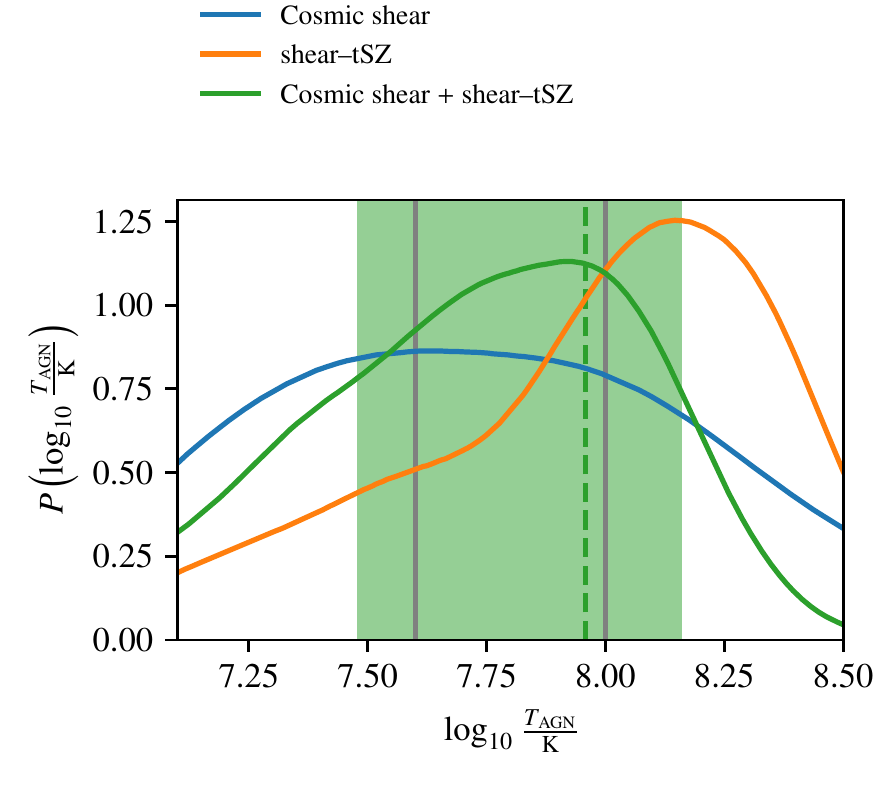}
                \caption{Marginal posterior densities of the baryon feedback strength, parameterised by \logtheatinline. 
                The green band and dashed line indicate the 68th percentile marginal CI and MAP, respectively, of the joint analysis of cosmic shear and the shear--tSZ cross-correlation. 
                The grey lines indicate the range of the feedback strength implemented in the \bahamas suite of hydrodynamical simulations, which our \hmx model is calibrated on. 
                \label{fig:logtheat}}
        \end{center}
\end{figure}
A key goal of this analysis has been to constrain baryonic feedback through the gas physics probed by the tSZ effect. 
In our model, the effect of baryonic feedback, specifically AGN feedback, is parameterised by the parameter \logtheatinline. 
The fiducial \bahamas model assumes $\logtheatinline=7.8$, since this value reproduces the observed present-day galaxy stellar mass function and the hot gas mass fractions of groups and clusters in the simulations \citep{McCarthy2017}. 
The constraints on this parameter, as inferred from cosmic shear alone, the shear--tSZ cross-correlation, and the joint analysis are shown in Fig.~\ref{fig:logtheat}. 
While the KiDS-1000 cosmic shear band-power and \pCl data by themselves have weak preference for low feedback strength \citep[][]{Asgari2021-CS, Troester2021}, the shear--tSZ cross-correlation data prefer strong feedback. 
For the joint analysis we find 
\begin{equation}
    \logtheat = \logtheatJoint\,,
\end{equation}
in good agreement with the \bahamas model. 
We note that this constraint does not strictly pass the criterion of \citet{Asgari2021-CS} for a parameter to count as constrained by the data: the marginal posterior density at the (lower) prior edge is not less than 13\,\% of the posterior density at the marginal mode. 

This result should be interpreted with caution, however. 
The model for the matter and electron pressure power spectra underlying our analysis, \hmx, was calibrated against the \bahamas suite of hydrodynamical simulations. 
The range of feedback strength considered in \bahamas ranges from $7.6$ to $8.0$, indicated by the grey band in Fig.~\ref{fig:logtheat}, and was chosen such that the simulations reproduce the observed galaxy stellar mass function and gas fraction in groups and clusters. 
Outside of this range the model extrapolates from the simulations it was calibrated on. 
While the physical nature of the model makes this extrapolation viable to a certain degree, there is significant modelling uncertainty at the edges of the $\logtheatinline\sim U(7.1, 8.5)$ prior we use in our inference pipeline. 
Furthermore, the fitting of the \hmx model parameters and their relation to \logtheatinline is done at a single, fixed cosmology. 
It has been shown that the effects of cosmology and galaxy formation are largely independent \citep[\eg][]{van-Daalen2011, Mummery2017, van-Daalen2020, Pfeifer2020}. 
Residual interplay between feedback parameters and cosmology, for example through the baryon fraction, can affect the inferred values of \logtheatinline to a small degree, however \citep{Mead2021-HMCode2020}.
To address these shortcomings, improvements in the modelling and its calibration will require suites of simulations that cover a wider range of cosmologies and feedback parameters. 
These simulations also need to cover larger volumes in order to suppress the sampling variance of the electron pressure power spectra at large scale (cf. Sect.~\ref{sec:3d-model} and \citealt{Mead2020}).

The shear--tSZ cross-correlations using the ACT maps do not constrain \logtheatinline. 
This can be attributed to the generally weaker constraints of the ACT maps due to the much smaller overlap with the KiDS-1000 data and the associated statistical scatter.

While the shear--tSZ cross-correlation by itself prefers strong feedback, the joint constraints on the feedback parameter agree with the \bahamas model. 
This shift in the preferred feedback strength is due to the interaction of IA and baryon feedback: both have the effect of a broadband suppression of the shear--tSZ cross-power spectra, such that high IA amplitudes can be compensated to a certain degree by weak baryon feedback.  
The joint analysis with cosmic shear restricts the range of allowed IA amplitudes, excluding the parts of parameter space with strong feedback. 
The IA model used in this analysis is quite simplistic, and the \hmx model is getting extrapolated to feedback strengths outside the range it is calibrated on. 
To better understand this IA--baryon feedback interaction, especially away from their fiducial parameters, a more holistic model is required, for example through including the halo-model IA formalism of \citet{Fortuna2021} in \hmx, as well as expanding the calibration set of hydrodynamical simulations.

One might have expected the shear--tSZ cross-correlation to yield tighter constraints on baryon feedback, since it is sensitive to both the effect of AGN feedback on gas physics and its suppression of the matter power spectrum. 
At large scales, $k\lesssim 0.3\iMpc$, stronger feedback increases the matter-electron pressure cross-power spectrum due to heating of the warm-hot intergalactic medium. 
These scales are largely outside of our scale cuts of $\ell\in(100,1500)$, however. 
At small scales, $k\gtrsim 0.3\iMpc$, increasing the feedback strength suppresses the matter-electron pressure cross-power spectrum \citep[see figure 9 in][]{Mead2020}. 
For the scales considered in this analysis, the effect of AGN feedback is thus largely restricted to a suppression of the angular cross-power spectrum over a large range of multipoles, similar to its effect on cosmic shear. 
This is illustrated in Fig.~\ref{fig:best-fit-model}, where the colouring of the samples of the TPD indicates the effect of \logtheatinline on the model prediction: except at scales of $\ell\sim100$, the effect of \logtheatinline is limited to a suppression of the angular power spectra. 
Conversely, ongoing and future large-area surveys, such as DES Y3, LSST, or \textit{Euclid}, have access to larger angular scales and thus could make use of the different behaviour at small and large scales to derive tighter constraints on baryon feedback.

\begin{table}
        \begin{center}
                \caption{Goodness of fit of our measurements.}
                \label{tab:goodness-of-fit}
               \begin{tabular}{lcccc}
        \toprule
        Data             & $\chi^2_{\rm MAP}$  & $N_\mathrm{data}$ &  $N_\mathrm{dof, eff}$ & PTE  \\
        \midrule
    
pseudo-$C_\ell$ cosmic shear& $150.7$ & $120$ & $117.1$ & $0.02$  \\
shear--tSZ& $28.0$ & $40$ & $36.2$ & $0.83$  \\
Cosmic shear + shear--tSZ& $179.3$ & $160$ & $155.4$ & $0.09$  \\

        \bottomrule
    \end{tabular}       \end{center}
        \tablefoot{We list the $\chi^2$ value at the maximum of the posterior, the size of the data vector, the effective number of degrees of freedom, and the         probability to exceed (PTE) the measured $\chi^{2}$ value.}
\end{table}

\section{Conclusion}
\label{sec:conclusions}
We conducted a \pCl cosmic shear analysis of the KiDS-1000 weak gravitational lensing data and find excellent agreement with previous KiDS-1000 angular power spectrum analyses \citep{Asgari2021-CS, Loureiro2021}. 
We then measured the cross-correlation of KiDS-1000 weak lensing and the tSZ effect, detecting the cross-correlation at a \SNTEMILCA significance. 

Using the \hmx model, which allows for the consistent modelling of both cosmic shear and the shear--tSZ cross-correlation -- including the effect of baryon feedback -- we derived constraints on cosmology and the baryon feedback parameter from the shear--tSZ cross-correlation. 
While the inferred value of the amplitude of the shear--tSZ signal, as parameterised by $\Sigma_8^{0.2}$, is fully consistent with other KiDS analyses, it is low by $2.2\,\sigma$ compared to CMB constraints from \Planck. 
Finally, we conducted a joint analysis of our cosmic shear and shear--tSZ cross-correlation data, taking the full cross-covariance between these data vectors into account. 
By breaking degeneracies, the inferred cosmological parameter constraints of the joint analysis are tighter than those of either probe by itself. 
For example, the constraints on $S_8$ of the joint analysis improve by \SeightJointImprovement over cosmic shear by itself. 
We find that our cross-correlation data prefer strong baryon feedback, while we infer $\logtheatinline = \logtheatJoint$ from the joint analysis, consistent with the \bahamas hydrodynamical simulations.  
Unlike recent work by \citet{Gatti2021} on shear--tSZ cross-correlation using DES Y3 data, we find no evidence of significant CIB contamination in our \Planck-derived Compton-\y maps.

While the analysis presented here demonstrates the power of shear--tSZ cross-correlations and their joint analyses with cosmic shear to constrain both cosmology and baryonic processes of galaxy formation, there are still significant challenges in the detailed interpretation of these results. 
The modelling of the non-linear matter spectrum to the accuracies required by current and future weak-lensing surveys is already a challenging proposition. 
Accounting for the electron-pressure field required to predict the tSZ effect adds a further level of complexity to the modelling framework. 
While the consistent modelling of various probes in \hmx is a significant improvement on earlier works that treat the profiles of different components as independent, its current accuracy for the matter--electron pressure power spectrum is limited at the 15 percent level and will need to be improved for future high-precision measurements, for example through the inclusion of non-linear halo biasing in the core halo model \citep{Mead2021b}. 
Improvements in the modelling must also go hand in hand with improvements of the simulation suites used for calibration, both in their coverage of the cosmology and baryon feedback parameter spaces and in their volumes to suppress sampling variance.

If these challenges can be overcome, the joint analysis of weak lensing from forthcoming galaxy surveys, such as LSST and \textit{Euclid}, with low-noise and high-resolution measurements of the tSZ effect from future CMB experiments, such as the Simons Observatory \citep{Ade2019-SO}, can be used to extract more information on galaxy formation processes and cosmology from these datasets than when analysed independently.
Another avenue would be to combine the measurements considered here with other datasets. 
The combination of weak lensing, the clustering of galaxies, and their cross-correlations into `$3\times 2$pt' analyses has provided some of the tightest constraints on cosmology from the low-redshift Universe  \citep[\eg][]{DES-Y1-3x2pt, Heymans2021, DES-Y3-3x2pt}. 
A lensing--tSZ $3\times 2$pt analysis, through inclusion of the tSZ auto-correlation in the measurements considered in this work, would thus be a logical next step and would build on previous work of joint analyses of the tSZ auto-power spectrum and lensing \citep{Osato2020, Makiya2020}. 
Beyond the shear--tSZ cross-correlation considered in this work, cross-correlations of galaxies with both the tSZ effect and the kinetic Sunyaev-Zeldovich effect have been shown to be a potent probe of the thermodynamics of haloes \citep[\eg][]{Battaglia2017, Schaan2021, Amodeo2021}. 
Adding these various Sunyaev-Zeldovich-effect cross-correlations into the framework of $3\times 2$pt analyses could therefore improve the constraining power of the data that are already available.

\begin{acknowledgements}
The figures in this work were created with \software{matplotlib} \citep{Hunter2007} and \software{getdist} \citep{Lewis2019}, making use of the 
\software{numpy} \citep{Oliphant2006} and \software{scipy} \citep{Jones2001} software packages. 
\\
This project has received significant funding from the European Union's Horizon 2020 research and innovation programme. 
We thank and acknowledge support from: 
the European Research Council under grant agreement No.~647112 (TT, AM, CH, and MA) and No.~770935 (HHi, AD, and AHW) in addition to the Marie Sk\l{}odowska-Curie grant agreements No.~797794 (TT) and No.~702971 (AM). 
We also acknowledge support from the Leverhulme Trust (TT);
the Max Planck Society and the Alexander von Humboldt Foundation in the framework of the Max Planck-Humboldt Research Award endowed by the Federal Ministry of Education and Research (CH);  
the University of British Columbia, Canada’s NSERC, and CIFAR (ZY and LVW); 
the Beecroft Trust and the Science and Technology Facilities Council through an Ernest Rutherford Fellowship, grant reference ST/P004474 (DA);  
the Polish Ministry of Science and Higher Education through grant DIR/WK/2018/12, and the Polish National Science Center through grants no. 2020/38/E/ST9/00395, 2018/30/E/ST9/00698, and 2018/31/G/ST9/03388 (MB); 
the Deutsche Forschungsgemeinschaft Heisenberg grant Hi 1495/5-1 (HHi);  
the Royal Society and Imperial College (KK); 
the NSFC of China under grant 11973070, the Shanghai Committee of Science and Technology grant No.19ZR1466600, the Key Research Program of Frontier Sciences, CAS, Grant No. ZDBS-LY-7013, and CMS-CSST-2021-A01 (HYS). 
\\
Based on data products from observations made with ESO Telescopes at the La Silla Paranal Observatory under programme IDs 177.A-3016, 177.A-3017 and 177.A-3018. 
\\
{ {\it Author contributions:}  All authors contributed to the development and writing of this paper.  The authorship list is given in three groups:  the lead authors (TT, AM, CH) followed by two alphabetical groups.  The first alphabetical group includes those who are key contributors to both the scientific analysis and the data products.  The second group covers those who have either made a significant contribution to the data products, or to the scientific analysis.}
\end{acknowledgements}

\bibliographystyle{aa}
\bibliography{references}

\begin{appendix}

\section{$B$-modes}
\label{sec:b-modes}
\begin{figure*}
        \begin{center}
                \includegraphics[width=\textwidth]{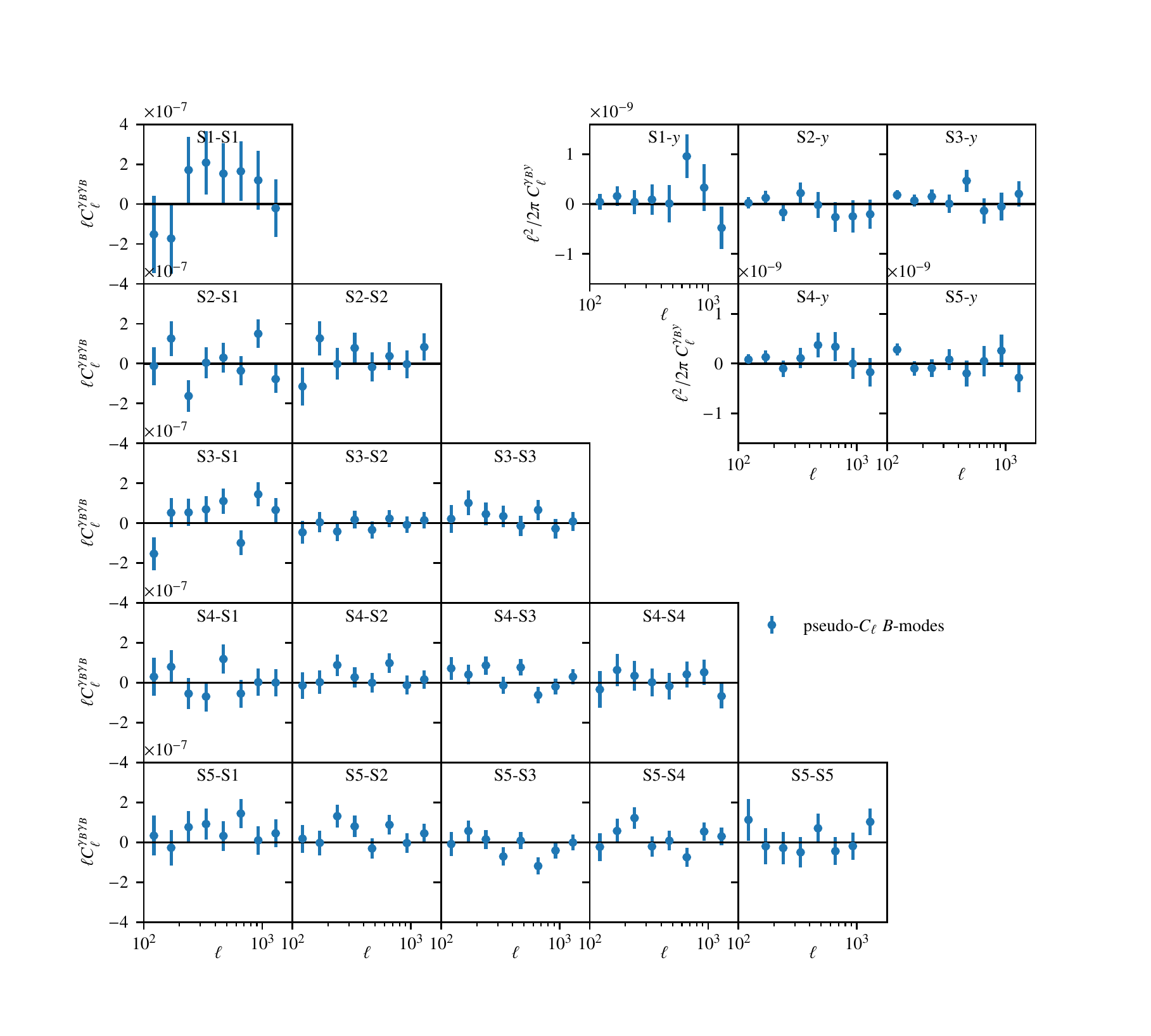}
                \caption{$B$-modes of the measurements considered in this work. 
                The cosmic shear $BB$ component is shown on the left, for which we find $\chi^2_{\gamma_B\gamma_B} = \chisqBB$, corresponding to a PTE  of \pteBB. 
                The $B$-mode of the shear--tSZ cross-correlation is shown in the upper right. 
                It has $\chi^2_{\gamma_B y} = \chisqTBMILCA$, for a PTE of \pteTBMILCA.
                \label{fig:b-modes}}
        \end{center}
\end{figure*}
Figure~\ref{fig:b-modes} presents the $B$-modes of our measurements, which are expected to be consistent with zero in the absence of systematics. 
To be more specific, for cosmic shear the $\gamma_B\gamma_B$ component of the angular power spectra is shown, while for the shear--tSZ cross-correlation, the $\gamma_B y$ component is depicted. 
Both \pCl measurements are consistent with zero, with $\chi^2_{\gamma_B\gamma_B} = \chisqBB$ and $\chi^2_{\gamma_B y} = \chisqTBMILCA$ for 120 and 40 degrees of freedom, respectively. 
This translates to probabilities to exceed (PTEs) of \pteBB and \pteTBMILCA, which makes us confident in the robustness of our measurements. 
The same is true for the other \y-maps considered in this work. 
Furthermore, the $B$-modes remain consistent with zero at the same level of significance when considering the whole $\ell\in(51, 2952)$ range they were measured on, rather than the $\ell\in(100, 1500)$ range used for parameter inference. 

\section{Comparison with previous KiDS-1000 cosmic shear results}
\label{app:bp-vs-pcl}
The cosmic shear analysis presented here closely followed that of \citet{Asgari2021-CS} but differs in two aspects: the choice of angular power spectrum estimator and the modelling of baryonic feedback. 

\begin{figure*}
        \begin{center}
                \includegraphics[width=\textwidth]{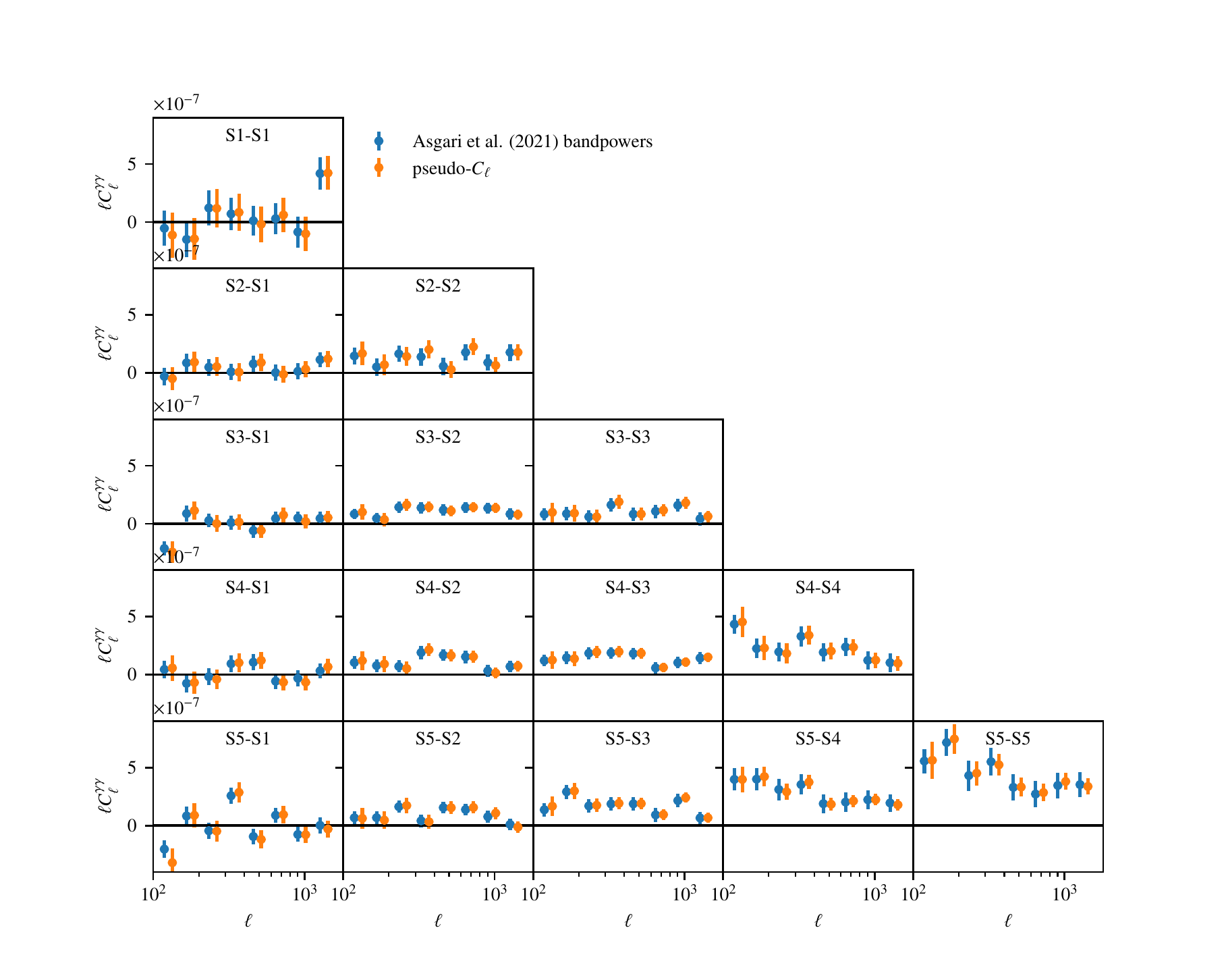}
                \caption{Comparison of KiDS-1000 angular power spectrum estimates. 
                The \pCl estimates used in this work are shown in orange, and the band-power estimates of \citet{Asgari2021-CS} are shown in blue.
                \label{fig:kids-nmt-measurements}}
        \end{center}
\end{figure*}

\begin{figure}
        \begin{center}
                \includegraphics[width=\columnwidth]{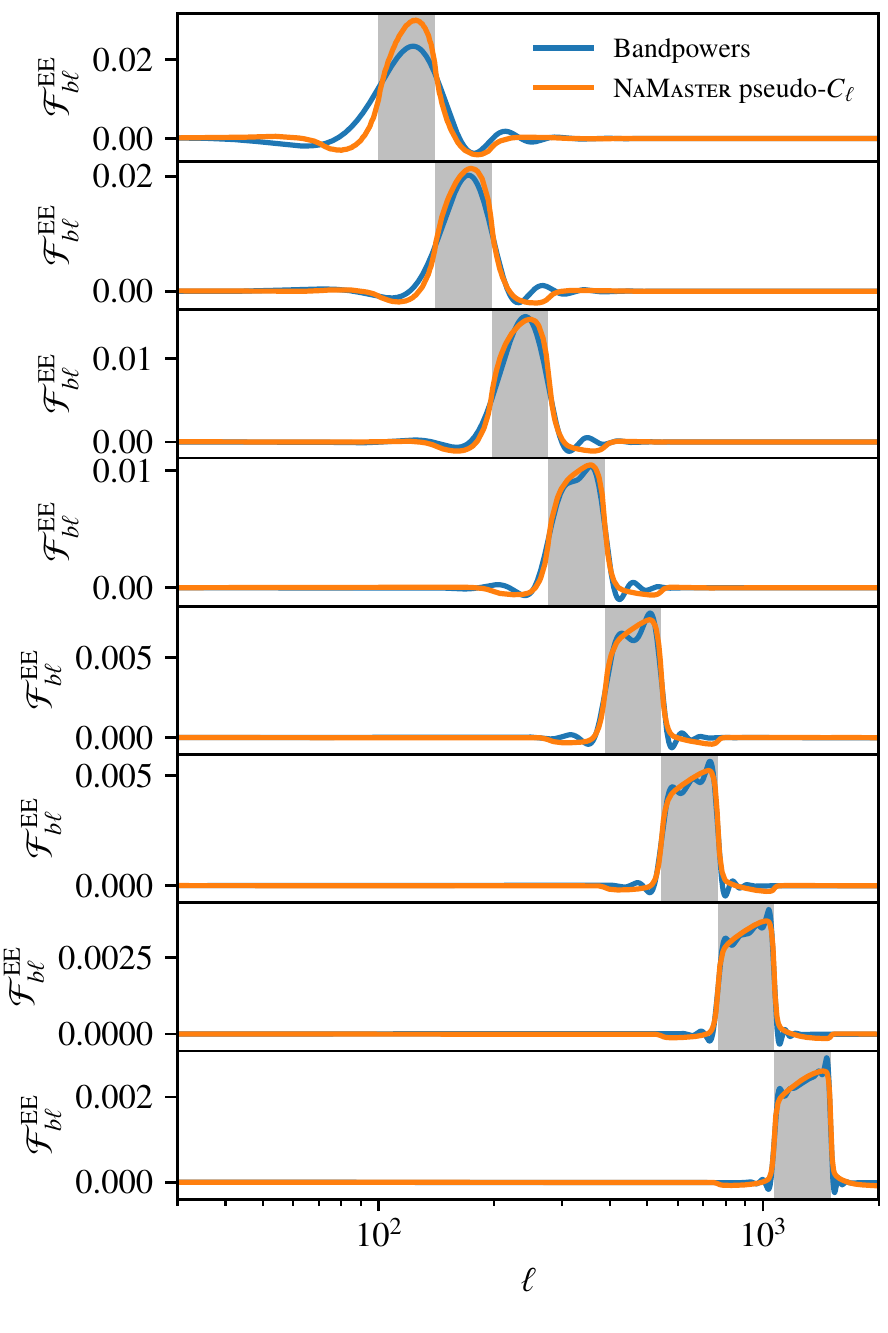}
                \caption{Window functions of Eq.~\eqref{equ:bandpower-window-functions}, relating the smooth angular power spectrum model to the binned prediction compatible with the measured data vector. 
                The differences between the multipole bin (grey bands) and the window function are due to masking and weighting of the shear maps in the case of the \pCl estimator (orange) and the finite angular range used for the band-power estimator (blue).
                \label{fig:bp-pcl-window-functions}}
        \end{center}
\end{figure}
\subsection{Comparison of the band-power and \pCl estimators}
Different two-point statistics estimators can give slightly different results \citep[\eg][]{HSC-Cl, HSC-xi, Asgari2021-CS}, as they are sensitive to different scales and are thus differently affected by statistical fluctuations in the underlying data. 
Here we compared our \pCl estimates and derived cosmological constraints to those from the band-power estimator used in \citet{Asgari2021-CS}. 
That band-power estimator is based on the integral transform of the two-point correlation function \citep[e.g.][]{Schneider2002, Becker2016a, Troester2017, van-Uitert2018, Joachimi2021} and has attractive properties, such as being insensitive to the footprint geometry nor requiring subtraction of the noise bias. 

We find excellent agreement between the measured data vectors, illustrated in Fig.~\ref{fig:kids-nmt-measurements}. 
Since both are estimators for the angular power spectrum, this is to be expected to a certain degree but it is reassuring the see this level of agreement. 
The uncertainties differ somewhat between the estimators, likely due to differences in the details of the data covariance calculation, for example, in the treatment of the mixed term of the Gaussian contribution. 
The window functions for the $EE$-component that translate the model prediction into data-space, accounting for example for residual effects of the mode-mixing matrices and $E/B$-mode mixing, are shown in Fig.~\ref{fig:bp-pcl-window-functions} for the two estimators. 
The largest differences are at large scales, where mask effects and the finite angular range on which the correlation function were measured become important. 

\begin{figure}
        \begin{center}
                \includegraphics[width=\columnwidth]{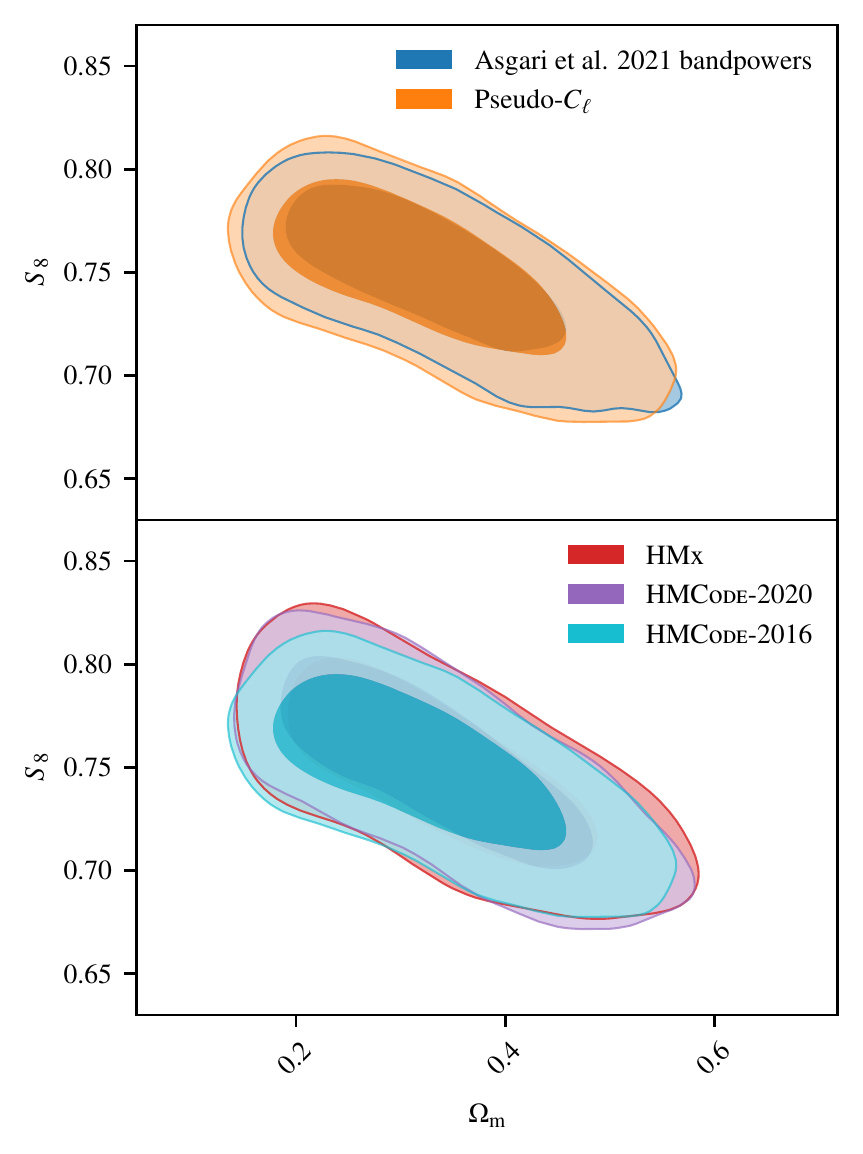}
                \caption{Comparison of the constraints on \om and $S_8$ between the band-power analysis of \citet{Asgari2021-CS} and the \pCl analysis of this work (top) and effect of different non-linear matter power spectrum modelling approaches on the cosmology constraints of the \pCl cosmic shear measurements (bottom). 
                \label{fig:bp-pcl-constraints}}
        \end{center}
\end{figure}

\begin{figure}
        \begin{center}
                \includegraphics[width=\columnwidth]{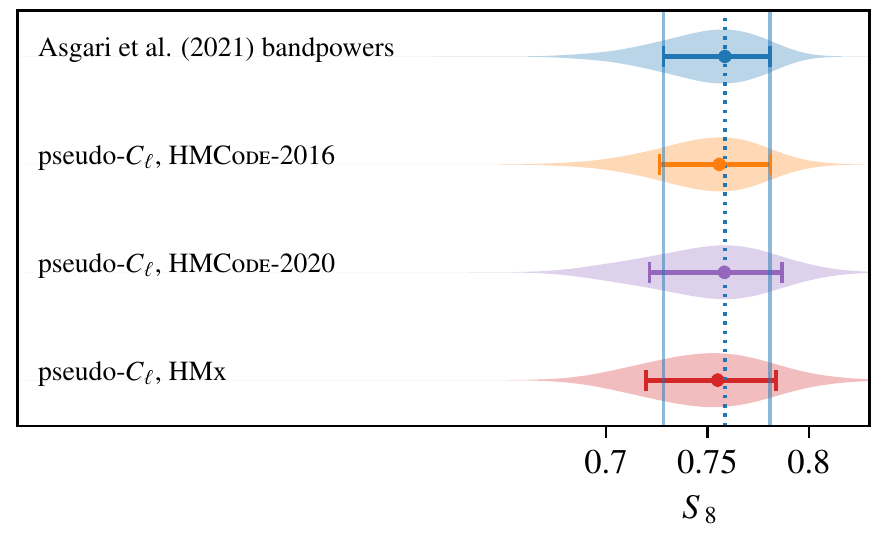}
                \caption{Marginal posterior densities of $S_8$, for the same data and models as in Fig.~\ref{fig:bp-pcl-constraints}. 
                The blue band and dashed line indicate the 68th percentile marginal CI and MAP of the \citet{Asgari2021-CS} band-power analysis, respectively.
                \label{fig:bp-pcl-constraints-S8}}
        \end{center}
\end{figure}

We derived cosmological constraints from both data vectors, taking care to homogenise the inference pipeline as much as possible. 
We use the same {\sc HMCode-2016} model \citep{Mead2016} for the matter power spectrum, as well as the multiplicative shear bias values from \citet{giblin/etal:2021} used in \citet{Asgari2021-CS}. 
The marginal posteriors in the \om-$S_8$ plane and for $S_8$ are shown in Figs.~\ref{fig:bp-pcl-constraints} and \ref{fig:bp-pcl-constraints-S8}, demonstrating good agreement between the two estimators. 
Due to differences in the way the data covariances for the two estimators were computed, small differences in the parameter constraints are to be expected. 

Our methodology differs in a subtler manner from the \pCl methodology of \citet{Loureiro2021}, however, even though both approaches make use of the \pCl formalism.  
In the present work the data vector was deconvolved with the mode-mixing matrices, the mode-mixing matrices were calculated using the weight map, and we used analytic estimates for the noise bias and data covariance. 
On the other hand, \citet{Loureiro2021} forward-modelled the effect of the mode mixing, used a binary mask to compute the mode-mixing matrices, and estimated their noise bias and data covariance from simulated mock data. 

\subsection{Impact of baryon modelling}
The second significant difference between the analysis choices of the present work and \citet{Asgari2021-CS} is in the non-linear model for the matter power spectrum. 
\citet{Asgari2021-CS} used {\sc HMCode-2016} \citep{Mead2016}, which was calibrated on the {\sc OWLS} suite of hydrodynamical simulations. 
The \hmx model used here has been calibrated against the \bahamas suite of simulations instead. 
Importantly, it models gas, stars, and dark matter separately. 
The model therefore does not have a dark-matter-only limit, in contrast to \hmcode-2016. 

Figures~\ref{fig:bp-pcl-constraints} and \ref{fig:bp-pcl-constraints-S8} show the effects of the different matter power spectrum model choices, comparing the \hmcode-2016 and \hmx models. 
The comparison also includes the {\sc HMCode-2020} \citep{Mead2021-HMCode2020} model, which has been optimised for the modelling of the matter power spectrum. 
Like \hmx, the baryonic feedback has been calibrated on \bahamas but unlike \hmx, it only predicts the matter power spectrum and thus cannot model the cross-correlations considered in the present analysis.
As discussed in appendix A of \citet{Troester2021}, {\sc HMCode-2020} allows for slightly higher values of $S_8$ because the finite amount of AGN feedback in the model always causes a degree of suppression of power. 
The same effect can be observed for \hmx, for the same reasons: the physical model for gas in the model prevents it from behaving as if it were dark matter.

\section{Other parameter constraints}
\begin{figure*}
        \begin{center}
                \includegraphics[width=\textwidth]{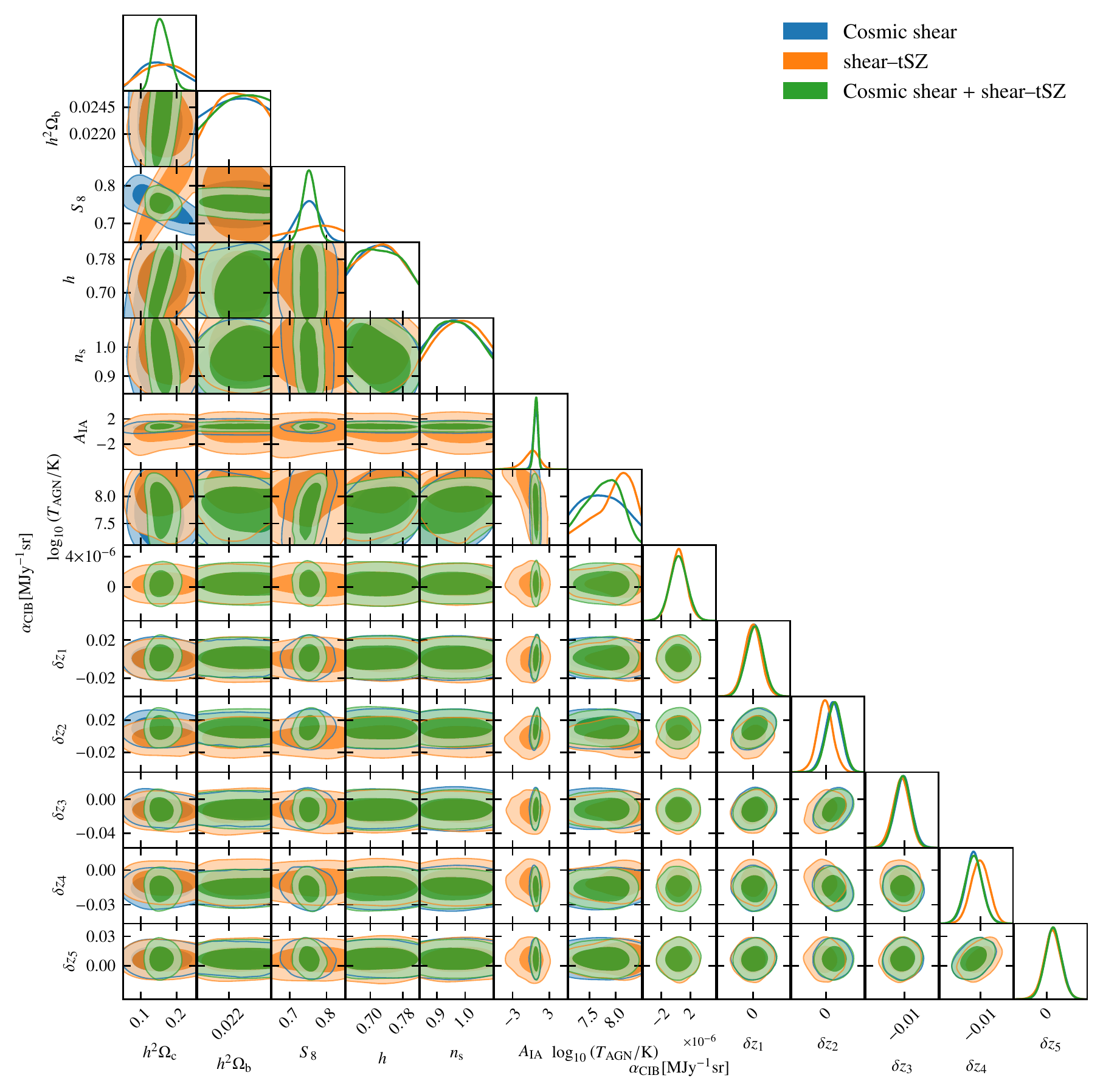}
                \caption{Constraints on all sampled parameters from KiDS-1000 cosmic shear, the cross-correlation of shear and the tSZ effect, and their joint analysis. 
                \label{fig:constraints-all}}
        \end{center}
\end{figure*}

Figure~\ref{fig:constraints-all} shows the marginal posteriors for all 13 sampled parameters in our fiducial cosmic shear, shear--tSZ, and joint analyses. 

\section{ACT measurements}
\label{app:ACT-maps}
\begin{figure}
        \begin{center}
                \includegraphics[width=\columnwidth]{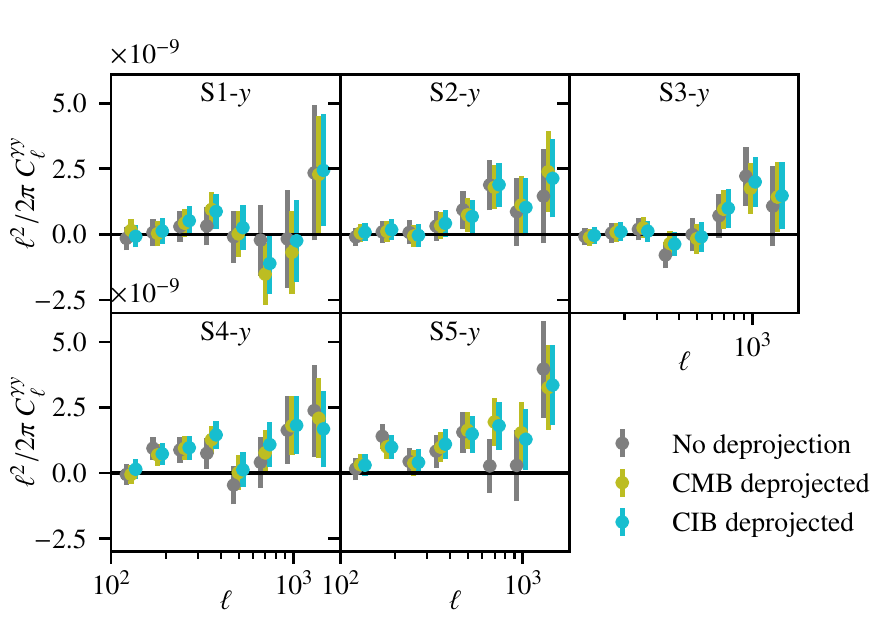}
                \caption{Measured angular cross-power spectra between KiDS-1000 shear and the ACT DR4 Compton-$y$ maps. 
                \label{fig:ACT-Cls}}
        \end{center}
\end{figure}

Figure~\ref{fig:ACT-Cls} shows the measured shear--tSZ angular cross-power spectra for different ACT Compton-$y$ maps provided by \citet{Madhavacheril2020}: 
\begin{description}
\item \texttt{tilec\_single\_tile\_BN\_comptony\_map\_v1.2.0\_joint}, corresponding to the case where no extra component has been deprojected;
\item \texttt{tilec\_single\_tile\_BN\_comptony\_deprojects\_cmb\_{\newline}map\_v1.2.0\_joint}, where the primary CMB has been deprojected;
\item \texttt{tilec\_single\_tile\_BN\_comptony\_deprojects\_cib\_{\newline}map\_v1.2.0\_joint}, where a CIB grey body spectrum has been deprojected. (See Sects.~\ref{sec:CIB} for further discussion of the CIB treatment in the ACT maps.)
\end{description}
The auto-power spectrum of the ACT $y$ map with no deprojection is higher than that of the CMB and CIB deprojected maps on the $\ell$-range we used in our analysis. 
This leads to 5-15\,\% larger errors on the shear--tSZ cross-correlation compared to the maps with deprojected components. 
The increase in the data covariance by itself does not explain the significantly weaker parameter constraints seen in Fig.~\ref{fig:constraints-CIB-banana}, however. 
Furthermore, the best-fit $\chi^2$ is higher than those of the maps with deprojected components, even though the data covariance is larger. 
Inspection of the measured \pCl, shown in Fig.~\ref{fig:ACT-Cls}, reveals that the highest redshift bin exhibits two data points that appear low, considering the strong correlation between the maps. 
To asses the impact of these data points, we analysed the shear--tSZ \pCl for the three ACT maps with the range $\ell \in (387.3, 1069.3)$ excised for the two highest redshift bins. 
This removes the tail to low values in the \Sigmaalpha posterior and brings all constraints into good agreement. 
Since we have no a priori reason to reject these data points, we refrain from further discussion of the constraints from these restricted data vectors. 
These findings warrant a more in-depth analysis of the ACT Compton-\y maps, which we leave for future work.

\section{\software{pyhmcode}}
We provide an easy to use and unified Python interface to \hmcode and \hmx, called \software{pyhmcode}\footnote{\url{https://github.com/tilmantroester/pyhmcode}}. 
A usage example is shown below. 
It assumes a \software{CCL} cosmology object \lstinline{ccl_cosmology} and arrays of the linear matter power spectrum \lstinline{pofk_lin}, redshift \lstinline{z}, and wave number \lstinline{k} (in units of $\iMpc$) and computes the auto- and cross-power spectra between the matter and electron-pressure fields.
\begin{lstlisting}[language=Python, basicstyle=\tiny\ttfamily]
import pyhmcode
from pyhmcode.halo_profile_utils import ccl2hmcode_cosmo

# Create the pyhmcode cosmology object. 
hmcode_cosmology = ccl2hmcode_cosmo(
    ccl_cosmo=ccl_cosmology,
    pofk_lin_k_h=k,
    pofk_lin_z=z,
    pofk_lin=pofk_lin,
    log10_T_heat=7.8)

# Create the halo model object, which holds
# information on the specific halo model
# to use. E.g., the HMCode or HMx version.
hmcode_model = pyhmcode.Halomodel(
    pyhmcode.HMx2020_matter_pressure_w_temp_scaling)

# Now we can compute the non-linear power spectrum,
# given the cosmology, halo model,
# and a list of fields.
hmcode_pofk = pyhmcode.calculate_nonlinear_power_spectrum(
    cosmology=hmcode_cosmology,
    halomodel=hmcode_model, 
    fields=[pyhmcode.field_matter,
            pyhmcode.field_electron_pressure])
\end{lstlisting}

\end{appendix}
\end{document}